\documentclass[12pt]{article}
\usepackage{amsmath, amsthm}
\usepackage{latexsym}
\usepackage{eufrak}
\newtheorem{Lem}{{\sc Lemma}}
\newtheorem{Th}{{\sc Theorem}}

\newtheorem{Cor}{{\sc Corollary}}
\newtheorem{Def}{{\sc Definition}}

\newcommand{\upstar}{\mbox{$^{\textstyle *}$}}
\newcommand{\lt}{\rightarrow}

\newcommand{\dts}{{,\ldots,\,}}
\newcommand{\bt}{{\bf t}}

\newfont{\shell}{msbm10 scaled\magstep1}

\def\R{\mbox{\shell R}}
\def\N{\mbox{\shell N}}
\def\Q{\mbox{\shell Q}}
\def\Z{\mbox{\shell Z}}

\newcommand{\cA}{\mathcal{A}}
\newcommand{\cC}{\mathcal{C}}

\newcommand{\cF}{\mathcal{F}}
\newcommand{\cG}{\mathcal{G}}
\newcommand{\cH}{\mathcal{H}}
\newcommand{\cI}{\mathcal{I}}
\newcommand{\cR}{\mathcal{R}}
\newcommand{\cS}{\mathcal{S}}
\newcommand{\cT}{\mathcal{T}}
\newcommand{\cX}{\mathcal{X}}

\newcommand{\coef}{\rm{coeff}}

\newfont{\smallshell}{msbm10 at 7pt}

\begin{document}

\title{Measured Multiseries and Integration}
\author{John Shackell}
\date{\today}
\maketitle
%\classno{26A12 (primary), 33F10, 68W30, 03C64 (secondary)}

\begin{abstract}
A paper by Bruno Salvy and the author introduced measured multiseries
and gave an algorithm to compute these for a large class of elementary
functions, modulo a zero-equivalence method for constants. This gave a
theoretical background for the Maple$^\copyright$  implementation that Salvy
was developing at that time.

The main result of the present article is an algorithm to calculate measured
multiseries for integrals of the form $\int h\sin G$, where $h$ and $G$ belong to
a Hardy field, $\cH$. The process can reiterated with the resulting algebra,
and also applied to solutions of a second order
differential equation of a particular form.
\end{abstract}

\section{Introduction}
The classical theory of asymptotic series has had a long and rich development. By contrast,
algorithms applicable to cases where power series are no longer sufficient are a much more
recent phenomenum.

There have been two main strands of development. The first rests on the theory of Hardy fields,
\cite{Sh_bk,Sh1,Richardsonetal}. The algorithms are function based with the output generally
in the form of multiseries expansions, although other forms have been tried.

Trigonometric functions whose arguments tend to infinity have long been used in expansions.
\cite{SaSh10} offered a theoretical basis for these which gave them precise meaning in terms
of the asymptotics of functions, and extended the methods of \cite{Richardsonetal} to such
cases. By using a result from \cite{vdHoeven97ii}, it was shown that these can be applied
to a large class of elementary functions, although restrictions are needed on trigonomentric
functions appearing inside exponentials or other trigonometric functions. This work provided
a theoretical underpining for the Maple$^\copyright$ implementation which Salvy was
developing at that time.

If $\cH$ denotes any Hardy field in which multiseries can be calculated,
the algorithm of \cite{SaSh10} can easily be extended to an algebra $\cH_\cG$,
generated by $\cH$ and a finite number of sines or cosines of its
elements. Our concern here is to give an algorithm for computing measured
multiseries when integration is added to the signature of $\cH_\cG$.\\

A second strand has been developed by van der Hoeven, based ultimately on the
theories of Ecalle, \cite{Ecalle4}. Here the prime objects of study are formal transseries
rather than functions; see \cite{vdH:ln} for example. This approach undoubtedly
has advantages. Greater generality can often be obtained and results
are sometimes easier to prove, for example with differential equations, \cite{vdH:ln}. 
However the functions are more remote and of course functions have meaning beyond their
asymptotic expansions. Also zero equivalence tends to be a larger issue with this approach.
Here we express the opinion that there is room for both approaches, and note that
\cite{vdH09} seeks to combine the benefits of both.\\

This paper makes use of a number of results from elsewhere. Firstly we need Theorem~2 of
\cite{SaSh10}, which itself makes substantial use of Theorem~7 from \cite{vdHoeven97ii}.
The latter represents the culmination of a classical line of research and provides an
important stepping stone in the continuing development of the asymptotics of elementary funtions.

Secondly we generalise the concept of an asymptotic field from \cite{Sh7}
to the measured form of multiseries.

\cite{Bosh4} contains some subtle and surprising results about Hardy fields and
trigonometric functions. We make technical use of some of the machinery there.\\

As in almost all areas of symbolic computation involving transcendental functions,
zero equivalence presents a problem in this paper. A variety of methods exist for
deciding equivalence of functions provided that this can be done for constants,
but for that we need a conjecture or the services of an oracle. All statements
we make about computability are subject to the caveat this brings.
See \cite{Richardson94a}, \cite{Sh_bk} and the references cited there. However
it is worth pointing out that mathematics has coped with this difficulty for a
large number of years. Well understood, tried and tested numerical routines are
in place for constants.\\

In Section~\ref{HFSection} we give a very brief introduction to
Hardy fields and collect a number of small results for later use.
Then in Section~\ref{MExpsSec} we look at measured expansions. The
definitions and basic properties of measured limits, coefficients
and measured multiseries from \cite{SaSh10} are presented. Some
slight differences arise in that \cite{SaSh10} treated asymptotics
at $x=0$, whereas we make the convention here that limits are at
$+\infty$. The obvious change of variable transforms between these.

Section~\ref{AsymValSec} looks at the algebra $\cH_\cG$ mentioned in the Abstract.
The (order reversed) valuation easily extends to this. Van der Hoeven considers a
more general case; \cite{vdH:ln}, \cite{vdH09}. Here we mainly confine ourselves
to the needs of Sections~\ref{AddIntsSec} and \ref{AsymFldInt}.
We establish a result concerning algebraic independence of sines.
Then a corresponding result is given relating to asymptotic
independence in a sense which is made precise. This is proved
for the quotient field of $\cH_\cG$ although the application made in
this paper needs only a weaker result.

Section~\ref{AsymFldsSec} is about asymptotic domains. The
corresponding concept in Hardy fields was introduced
in \cite{Sh7} in order to pass from an algorithm in a
Hardy field, $\cF$, to an algorithm in a suitable extention,
$\cF(f)$. In particular the case when $f$ is an integral of an
element of $\cF$ was treated there. Here we extend these ideas
to our `measured' context.

In Section~\ref{AddIntsSec}, we consider the integrals
of the form $\int h\sin G$ with $h,G\in\cH$ and $G\lt\infty$.
Some results from \cite{Bosh4}, are used to prove that
these have the form $f_1\sin G+f_2\cos G$ with $f_1$ and $f_2$
belonging to a Hardy field extension of $\cH$. As a corollary we obtain
that if $h_2=o(h_1)$ then $\int h_2\sin G$ and $\int h_2\cos G$ are
$o(\int h_1\sin G)$. It then follows that partial expansions of
$\int h\sin G$ generated using integration by parts do indeed
give asymptotic expansions. However these
are insufficient for our purpose since they are not multiseries.

The heart of the paper is Section~\ref{AsymFldInt}. There we give the
breakdown into cases for `the next term' of the multiseries, and use
the machinery developed in the previous sections to prove that the requisite
conditions are satisfied. It then follows that measured
multiseries can be computed in $\cH_\cG(\int\cH_\cG)$. Moreover it turns out
that we can integrate again and repeat the process.

Finally we show how our results can be used to extend
asymptotic fields by solutions of differential equations of the form
$gy^{\prime\prime}-g^\prime y^\prime+g^3y=f$, where $g,f\in\cH$.
\section{Hardy fields}\label{HFSection}
These were first defined in \cite{Bourbaki61}, and the
theory developed in \cite{Robinson72}, \cite{Bosh1,Bosh2} and
\cite{Rosenlicht83,Rosenlicht83ii}, and by other authors.

Let $\chi$ denote the ring of germs of real-valued functions at $+\infty$.
\begin{Def}
A {\em Hardy field} (at $+\infty$) is a differential subfield of $\chi$.
\end{Def}
It is not hard to show that a Hardy field carries a total order which
reflects the asymptotic behaviour of the elements; so $f>g$ if this is
true for sufficiently large values of $x$. Then $f^\prime(x)$ has constant
sign for $x$ large, and it follows that any element
of a Hardy field tends to a limit, finite or infinite.

Some notation should be mentioned here. For $f$ a Hardy-field
element, $\log f$ will mean $\log |f|$ and $\log_k$ will denote the $k$-times
iterated logarithm. We write $f^\Delta$ for the logarithmic derivative
of $f$; i.e. $f^\prime/f$.

It is possible to extend a Hardy field in various ways and obtain another
Hardy field. For example one can add exponentials, logarithms and more
generally integrals of existing elements; see for example
 \cite{Rosenlicht83}. Note however that it is not generally true that
the union of two Hardy fields is contained in a Hardy field,
\cite{Bosh2,Rosenlicht83ii,Bosh4}.

Sometimes one wants a coarser comparison than the Hardy-field ordering
gives. If $\cH$ is a Hardy field and $f$ and $g$ are any two non-zero
elements of $\cH$, we define $f\asymp g$ to mean that $f/g$ tends to a
finite non-zero limit, c.f. \cite{Hardy12}.
Then $\asymp$ is an equivalence relation, and we
write $\gamma_0(f)$ for the equivalence class of $f$. Addition of the
equivalence classes is obtained by setting
$\gamma_0(f)+\gamma_0(g)=\gamma_0(fg)$; this is easily seen to be
well defined.

It is convenient to order these classes with the ordering induced from
$\cH$. Then $\gamma_0$ is just the valuation with the reverse of the
usual ordering; see \cite{Rosenlicht83ii}. We shall often write $f\succ  g$ to
indicate that $\gamma_0(f)>\gamma_0(g)$.

Now suppose that $f$ and $g$ are two elements of $\cH$ which tend to
infinity. We put $f\sim_1g$ if there is a positive integer $n$ such that
$f^n\succ g$ and $g^n\succ f$. We specify
that $\pm f^{\pm 1}$ are all related to each other under $\sim_1$, and that
all elements tending to a non-zero constant are likewise related.
Then $\sim_1$ is again an equivalence relation, and we write $\gamma(f)$
for the resulting {\em comparability class} of $f$, \cite{Rosenlicht83ii}.
The comparability classes are ordered by taking $\gamma(f)>\gamma(g)$ when
$\log |f|\succ \log |g|$, or alternatively for $f,g\lt\infty$, when $f>g^n$
for all $n\in\N$.

The class $\gamma$ is particularly important for multiseries expansions
because elements of different comparability class are mutually
transcendental and do not intefere asymptotically with one another.
It is possible to define $\gamma_2$ and higher gammas.
In particular, $\gamma_2$ sometimes plays a role in asymptotics,
\cite{Sh_bk}. However we shall not use it here.

A very useful notation denotes that two elements of $\cH$ are
asymptotic at a level which sits between $\gamma_0$ and $\gamma$.
For $f,g$ elements of $\cH$ with $f,g\not\asymp 1$
we write $f\bowtie g$ to indicate that
$\gamma(f)=\gamma(g)>\gamma(f/g)$; i.e. $f=g^{1+o(1)}$. It is
convenient to also specify that $f\bowtie g$ whenever
$f,g\asymp 1$.

When working with Hardy fields one usually needs a number of small
results, \cite{Bosh1,Bosh2,Rosenlicht83,Rosenlicht83ii,Sh_bk}.
Some of these are collected together in the following lemma.
\renewcommand{\theenumi}{(\roman{enumi})}
\begin{Lem}     \label{HFsmalls_lem}
  Let $h,h_1,h_2$ belong to a Hardy field $\cH$. Then the following hold:
\begin{enumerate}
\item If $h\lt K\in\R$ then $h^\prime\lt 0$. If $h_1\sim h_2$ and
$1\not\asymp h_1\;(\asymp h_2))$ then $h_1^\prime\sim h_2^\prime$.
\item Suppose that $h_1\not\asymp 1$. Then $h_1\asymp h_2$
implies $h_1^\prime\asymp h_2^\prime$, and $h_1\succ h_2$ implies
$h_1^\prime\succ h_2^\prime$.
\item If $h_1,h_2\not\asymp 1$, then
$h_1^\Delta\asymp h_2^\Delta$ if and only if $\gamma(h_1)=\gamma(h_2)$,
and $h_1^\Delta \succ  h_2^\Delta$ if and only if $\gamma(h_1)>\gamma(h_2)$.
\item If $\gamma(h_1)>\gamma(h_2)$ then $(h_1h_2)^\prime\sim h_1^\prime h_2$.
\item If $\gamma(h)>\gamma(x)$ then $\gamma(h^\Delta) <\gamma(h)$
(i.e. $h^\prime\bowtie h$).
\item If $\gamma(h)<\gamma(x)$ and $h\not\asymp 1$, then
$\log h^\prime\sim -\log x$, and so $h^\prime\bowtie x^{-1}$.
\item If $\gamma(h)=\gamma(x)$ and $\log h\not\sim\log x$ then
$\gamma(h^\prime)=\gamma(x)=\gamma(h)$.
\item If neither of $h_1$ and $h_2$ is asymptotic to a non-zero constant,
then $h_1^\prime /h_2^\prime\bowtie h_1/h_2$.
\end{enumerate}
\end{Lem}
With reference to (v), note that the conclusion may fail without the
restriction $\gamma(h)>\gamma(x)$. An example is given by $h(x)=x^{-1}$.

\par\begin{proof}%[of Lemma~\ref{HFsmalls_lem}]
Most of these results have been proved elsewhere, for example in \cite{Sh_bk},
and are in any case not hard. We therefore give proofs for (v) and (viii) only.

To prove (v) we may suppose that $|\,h|\lt\infty$; for otherwise we may replace
$h$ by $h^{-1}$. Then $\log |\,h|>K\log x$ for every $K\in\R$. On differentiating,
we obtain that $h^\prime/h>K/x$, and it follows that $h^\prime\succ h^{1-o(1)}$.
On the other hand, $h^{-d}\lt 0$ for $d\in\R^+$ and hence $h^\prime/h^{1+d}\lt 0$.
Since the hypotheses ensure that $|\,h^\prime |\lt\infty$, this implies that
$h^\prime\prec h^{1+d}$. Thus $h^\prime\bowtie h$.

As regards (viii), if $\gamma(h_1)=\gamma(h_2)$ then (iii) gives
$h_1^\Delta\asymp h_2^\Delta$ and hence $h_1^\prime/h_2^\prime\asymp
h_1/h_2$. So we may now assume that  $\gamma(h_1)>\gamma(h_2)$.

If $\gamma(h_1)>\gamma(x)$ then by (v),
$h_1^\prime\bowtie h_1$ and $\gamma(h_1)>\gamma(h_2^\prime)$. Hence
$h_1^\prime/h_2^\prime\bowtie h_1\bowtie h_1/h_2$.

For the case when $\gamma(h_1)\leq\gamma(x)$, suppose first that
$h_1\lt\infty$. We write $\lambda=h_1^{inv}\circ\exp$, so that
$h_1\circ\lambda=e^x$. By the case already proved,
\[ \frac{(h_1\circ \lambda)^\prime}{(h_2\circ \lambda)^\prime}
\;\;\bowtie\;\;\frac{h_1\circ \lambda}{h_2\circ \lambda}. \]
The relation $\bowtie$ is unaffected by composition with
a function tending to infinity. Therefore
\[ \frac{h_1^\prime}{h_2^\prime}=
\frac{(h_1\circ \lambda\circ \lambda^{inv})^\prime}
{(h_2\circ \lambda\circ \lambda^{inv})^\prime}=
\frac{(h_1\circ \lambda)^\prime\circ \lambda^{inv}}
{(h_2\circ \lambda)^\prime\circ \lambda^{inv}}
\bowtie\frac{(h_1\circ \lambda)\circ \lambda^{inv}}
{(h_2\circ \lambda)\circ \lambda^{inv}}=
\frac{h_1}{h_2}. \]
If $h_1\lt -\infty$, we replace $h_1$ by $-h_1$, while if $h_1\lt 0$ we replace
$h_1$ by $1/h_1$ and $h_2$ by $1/h_2$.
\end{proof}

When $\cH$ is a Hardy field, we write $\cH\upstar$ for the set of non-zero
elements of $\cH$ and $\cH^+$ for the set of positive elements of $\cH$.
We shall use the corresponding notation for other structures.
\section{Measured expansions}\label{MExpsSec}
\subsection{Measured limits}
The frequent occurrence of sines and cosines in the expansions that arise in
applications makes it highly desirable to extend the theory of
multiseries, \cite{Richardsonetal,Sh_bk}, to allow their inclusion. This
problem was addressed in \cite{SaSh10}; a more primitive version was
given in \cite{SaSh02} and \cite{Sh_bk}. The basic
idea is to allow subsets of the range of arbitrarily small relative size
to be discounted for the asymptotics. We summarise here the main
definitions and results from \cite{SaSh10}, transformed so that limits
are at $+\infty$.
\begin{Def}     \label{limdef}
Let $f$ be a function from $\R$ to $\overline{\R}=_{def}\R\cup\{-\infty,\infty\}$.
Let $l\in\R$ and $\alpha\in\cH^+$ with $\alpha\lt 0$.
We say that $f$ has {\em measured limit} $l$ with respect to $\alpha$ as
$x\lt\infty$, and write $f\lt_\alpha l$, if for all $\varepsilon\in\R^+$
\[\frac{1}{\alpha(X)}\int_X^\infty 1_{|f(x)-l|>\varepsilon}d\alpha \lt 0 \]
as $X\lt\infty$. Similarly we write $f\lt_\alpha\pm\infty$ if, as $X\lt\infty$,
\[ \frac{1}{\alpha(X)}\int_X^\infty1_{f(x)<1/\varepsilon}d\alpha\lt 0
\qquad\mbox{respectively}
\qquad\frac{1}{\alpha(X)}\int_X^\infty1_{f(x)>-1/\varepsilon}d\alpha\lt 0. \]
\end{Def}
If $f$ tends to $l$ in the ordinary sense then $f\lt_\alpha l$ for any
$\alpha\lt 0^+$. The usual arithmetic properties of limits are also obtained for
measured limits, and one has the expected result for composition on the left;
i.e. if $f\lt_\alpha l$ and $g$ is continuous at $l$ then $g\circ f\lt_\alpha g(l)$.

Composition on the right needs more care. For example
$x^{-1}\sec x\lt_{x^{-1}}0$ despite the poles of sec, but if we compose
on the right by $\log_2$ the measured limit with respect
to $x^{-1}$ no longer exists. The oscillations of the secant are now so slow
that a non-negligible contribution to the integral builds up over a cycle,
\cite{SaSh02}. This behaviour is undesirable, and the answer is to use a
different $\alpha$, here $\log_2^{-1}x$. In general, if $f\lt_\alpha l$ and
$g\in\cH\lt\infty$ then $f\circ g(x)\lt_{\alpha\circ g} l$.
The price of introducing $\alpha$ is extra complication.
However the following result from \cite{SaSh10} provides some mitigation.
\begin{Lem}\label{scalelem1}
Let $\alpha$ and $\beta$ be two elements of $\cH$ which tend to zero, and
let $f$ be a function from $\R$ to $\overline{\R}$ such that $f\lt_\alpha l$
with $l\in\overline{\R}$. If $\gamma(\beta)\leq\gamma(\alpha)$ then
$f\lt_\beta l$.
\end{Lem}
The following definition now makes sense.
\begin{Def}\label{scaleHdef}
Let $f$ and $l$ be as above. We write $f\lt_\cH l$ if there exists an
$\alpha\in\cH^+$ with $\alpha\lt 0$ such that $f\lt_\alpha l$.
\end{Def}
\subsection{Coefficients}\label{CoeffSubSec}
Heuristically, a function has measured limit $l$ if it tends to $l$ on `most
of the range'. Our non-constant coefficients go to the opposite extreme;
they are very rarely near any particular value `in the limit'. The formal
definition from \cite{SaSh10} is as follows.
\begin{Def}\label{coeffdef}
Let $\alpha\in\cH^+$ tend to zero.
An element of $\coef_\alpha$ is either a real constant or is a function from
$\R$ to $\overline{\R}$ with the following property. For all
$l\in\R$, $\varepsilon\in\R^+$, there exist $\delta=\delta(l,\varepsilon)$
and $X_0=X_0(l,\varepsilon)\in\R^+$ such that for all $X>X_0$,
\begin{equation}        \label{coeffdefeq1}
\frac{1}{\alpha(X)}\int_X^\infty 1_{|f(x)-l|<\delta}\:d\alpha<\varepsilon\quad
\quad\mbox{and}\quad\quad \frac{1}{\alpha(X)}\int_X^\infty 1_{|f(x)|>\delta^{-1}}
\:d\alpha<\varepsilon.
\end{equation}
\end{Def}
We do not require the left-hand sides of (\ref{coeffdefeq1})
to tend to zero as $X\lt\infty$ for fixed $\delta$. If we did then $\sin x$
would fail to qualify as a coefficient!

The basic properties of coefficients, proved in \cite{SaSh10}, are as follows.
\begin{Lem}\label{coefflem}
Let $c$ be a non-constant element of $\coef_\alpha$, let $k\in\R$ and
let $h\lt_\alpha 0$.
\begin{enumerate}
\item The elements $c^{-1}$, $c+k$, $ck$ and $c+h$ are all in $\coef_\alpha$.
\item $ch\lt_\alpha 0$.
\item If $\rho\in\cH$ and $\rho\lt\infty$, then
$c\circ\rho\in\coef_{\alpha\circ\rho}$.
\item Let $\beta\in\cH^+$ tend to zero, and suppose that $\gamma(\beta)\leq
\gamma(\alpha)$. Then $\coef_\beta\supseteq\coef_\alpha$.
\end{enumerate}
\end{Lem}
A consequence of (iv) is that we can sensibly define $\coef_\cH$ as
$\cup\{\coef_\alpha; \alpha\in\cH^+\: \&\: \alpha\lt 0\}$.
One apparent problem is that $\coef_\alpha$ is not closed under arithmetic
operations. However in practice we consider subfields of $\coef_\alpha$,
or of $\coef_\cH$.
Theorem~2 of \cite{SaSh10} shows that many functions built from trigonometric
functions are in a suitable coefficient field. In particular all rational
functions of the elements of $S_\cT$ of Section~\ref{AsymValSec} are in
$\coef_\cH$.

We note that it is possible for a non-zero $\cC^\infty$ element of $\coef_\cH$
to vanish on a sequence of intervals of positive length tending to infinity
if the intervals are sufficiently spaced out along the real line.
\subsection{Measured multiseries}
The output from a calculation in asymptotics needs to be presented with the most
important information first. This idea leads naturally to the classical concept
of an {\em asymptotic series}, \cite{deBruijn58}. The following `measured' version
is from \cite{SaSh10}.
\begin{Def}\label{alphaasymdef}
Let $f_n(x),\; n\in\N$ and $F(x)$ be functions defined on an interval of
the form $(a,\infty)\subset\R$, except possibly on a set of measure zero.
Suppose that $\alpha\in\cH^+$ with $\alpha\lt 0$ and that for each $n\in\N$,
$f_{n+1}(x)/f_n(x)\lt_\alpha 0$. If also
\[ f_N^{-1}(x)\left (F(x)-\sum_{n=0}^Nf_n(x)\right )\lt_\alpha 0 \]
for each $N\in\N$, we  write $F(x)\sim_\alpha\sum f_n(x)$ and say that the series
$\sum f_n(x)$ is \emph{$\alpha$-asymptotic} to $F(x)$.
\end{Def}
Of course the series $\sum f_n(x)$ does not have to
converge. It will normally do so when attention is restricted to elementary
functions as in \cite{SaSh10}, but classical examples show that this need
no longer be the case when integration is introduced into the signature,
\cite{deBruijn58}.

The first asymptotic series studied were power series, but we also need to
use powers of logarithms, exponentials and more complicated objects in our
expansions. The scale elements that we use in this way should be of pairwise
different comparability class to avoid problems of cancellation.

Expansions sometimes need to take place inside exponentials, which may
result in a  new scale element. So the definitions of scale and expansion
mutually recurse, \cite{SaSh10} and c.f. \cite{Ecalle4,vdH:ln}.
\begin{Def}\label{scaledef}
An \emph{asymptotic scale} is a finite ordered set, $\{\bt_1\dts\bt_m\}$,
of positive elements of a given Hardy field which tend to
zero and satisfy the following conditions:
\begin{enumerate}
\item $\log\bt_i/\log\bt_{i+1}\lt 0$, for~$i=1,\dots,m-1$.
\item Each $\bt_i$ is either of the form $\log_k^{-1}x$, $k\geq 0$, or else $\log\bt_i$
has a multiseries expansion in the scale $\{\bt_1\dts\bt_{i-1}\}$.
\item $x^{-1}\in\{\bt_1\dts\bt_m\}$, and if $\log_k^{-1}x$ belongs to
$\{\bt_1\dts\bt_m\}$ for some $k>0$, then so do $\log^{-1}x$, \ldots,
$\log_{k-1}^{-1}x$.
\end{enumerate}
\end{Def}
Note that since measured limits are equivalent to ordinary limits in a Hardy
field, no extra generality is obtained by making the limits and multiseries in
the definition measured ones, as was done in \cite{SaSh10}. Condition (iii)
is innocuous and is included for technical reasons which need not concern
us here.

We may wish to consider non-integral constant powers of scale elements.
In that case the exponents need to belong to a finitely generated subset
of $\R$; i.e. a set of the form
$\lambda_1\N+\lambda_2\N+\dots+\lambda_k\N+\zeta$,
with $\zeta\in\R$ and $\lambda_1\dts\lambda_k\in\R^+$.
\begin{Def}\label{t1mmsdef}
Let $f$ be a real-valued function defined on an interval $(a,\infty)\in\R$
and let $\{\bt_1\dts\bt_m\}$ be an asymptotic scale.
Let $r_i$, $i=0\dts\infty$ be a sequence of elements of a finitely generated
subset of $\R$ with $r_i$ strictly increasing to infinity, and let
$c_m\in\coef_{\bt_1},m=0\dts\infty$. We say that $f$ has a $\bt_1$
{\em measured multiseries expansion} $\sum c_i\bt_1^{r_i}$
if the series is $\bt_1$-asymptotic to $f(x)$ in the sense of
Definition~\ref{alphaasymdef}.
\end{Def}
\begin{Def}\label{tmmmsdef}
Let $f$ and $r_i,i=0\dts\infty$ be as above and suppose $m>1$.
A $\{\bt_1\dts\bt_m\}$ measured multiseries expansion of $f$ is a series
$\sum g_i\bt_m^{r_i}$ which is $\bt_m$-asymptotic to $f(x)$, and where each
$g_i$ has a $\{\bt_1\dts\bt_{m-1}\}$ measured multiseries expansion.
\end{Def}

The main algorithm of \cite{SaSh10} shows how to compute measured multiseries
for a large class of elementary functions, although one has to avoid any
application of exponential or trigonometric functions to expressions of
the type $h\sin G$ where $h,G\lt\pm\infty$.\\

We denote the coefficient of $\bt_{m-1}^j$ in the $\bt_{m-1}$ expansion of $g_{i_1}$
by $g_{i_1,j}$, and then the coefficient of $\bt_{m-2}^k$ in the $\bt_{m-2}$
expansion of $g_{i_1,i_2}$ by $g_{i_1,i_2,k}$, and so on. To say that we can compute
the above multiseries for $f$ means that we must be able to specify each of
these coefficients in a form to which zero-equivalence methods may be applied.

For a particular $g_{i_1\dts i_s}$ we may assume that $g_{i_1\dts i_{s-1},0},\dts g_{i_1\dts i_{s-1},i_s-1}$
have already been specified in a suitable form. By subtracting these from $f$
and dividing by the leading power of $\bt_{m-s}$, we reduce to the case when $i_s=0$.
So our task is to say what should be the term in $\bt_{m-s}^0$ of the expansion of $f$ in powers
of $\bt_{m-s}$. To simplify notation in what follows, we replace $m-s$ by $i$ and call
the constant term the $i$-th {\em shadow} of $f$; we denote it by $\eta_i(f)$.

For an integral it is not at all obvious what we should take as the shadow. One property
we shall require of our method of choosing shadows is that algebraic combinations $\eta_i(f)$
for different $f$ shall not contain a positive power of $\bt_i$. We return to this matter
in Section~\ref{AsymFldsSec}.

\section{The algebra $\cH_\cG$}\label{AsymValSec}
Let $\cH$ be a Hardy field at $+\infty$, and suppose that there is a scale
$\{\bt_1\dts\bt_m\}$ in $\cH$ with respect to which every element of $\cH$ has
a multiseries. Let $\cG$ be a finite set of elements of $\cH$ each tending to
infinity. We are interested in the differential algebra, $\cH_\cG$, generated
by $\cH$ and the sines and cosines of the elements of $\cG$.

Any $G\in\cG$ may be written in the form $G=G_\infty+G_c+G_0$ where
$G_\infty$ contains those terms in the multiseries
which tend to $\pm\infty$, $G_c$ is the constant term and $G_0$ contains
the multiseries terms which tend to zero; see \cite{SaSh10} for example.
The functions $\sin G_c$, $\cos G_c$, $\sin G_0$ and $\cos G_0$ may be
added to $\cH$ as in \cite{Sh9}; see also \cite{Bosh1}. So we may take the
arguments of sines and cosines of elements of $\cG$ to have all multiseries
terms tending to infinity.
Further we may use the addition formulae to  arrange
that the elements of $\cG$ are linearly independent over $\Q$ and
are constant multiples of a subset of $\cG$ in which no two elements have the
same $\gamma_0$. Again we shall assume that this has been done.

We may write any $a\in\cH_\cG$ in the form
\begin{equation}\label{rvalsumform}
  a=\sum_1^m h_i\theta_i
\end{equation}
where $h_1, h_2\dts h_m\in\cH\upstar$ with $h_i\succ h_j$ for $i<j$ and
$\theta_1\dts \theta_m\in\R[\cG]$. We then extend $\gamma_0$ to $\cH_\cG$ by writing
$\gamma_0(a)=\gamma_0(h_1)$ and using the inherited addition and ordering on
the $\gamma_0$ classes.

We note that the methods of \cite{SaSh10} can be used to obtain multiseries
in $\cH_\cG$ provided this can be done in $\cH$.

For what follows we want to write a typical element of $\cH_\cG$ in a form slightly
different from (\ref{rvalsumform}).
Suppose that $G_\nu,\:\nu=1\dts u$ are elements of $\cG$ such that
$G_{\nu-1}\prec  G_\nu$ for $2\leq \nu\leq u$. Let $n_1\dts n_u\in\N^+$
with $n=n_1+\dots+n_u$, and for each $\mu=1\dts u$, suppose that
$\lambda_{\mu,1}\dts\lambda_{\mu,n_\mu}$ are positive real numbers linearly
independent over $\Q$. Set
\begin{equation}\label{STDefn}
S_\cT=\{\sin(\lambda_{1,1}G_1(x))\dts\cos(\lambda_{1,n_1}G_1(x))
\dts\cos(\lambda_{u,n_u}G_u(x))\}.
\end{equation}
A typical element of $\cH_\cG$ may be written in the form
\begin{equation}\label{TypicalElement}
P=\sum_jp_j\sigma_j,
\end{equation}
where the $p_j\in\cH$ and the $\sigma_j$ are pairwise distinct power
products of the various elements of $S_\cT$ having no square or higher
power of a cosine. It is possible that $\sigma_j=1$ for one $j$,
this being given by the empty product.

The following will be needed later in this section and also in
Section~\ref{shadowSFii}.
\begin{Lem}\label{zerolts_lem}
Let $P\in\cH_\cG$. Suppose that there is a sequence $\{z_n\}$ of points
tending to infinity, at each of which $P$ and all its derivatives vanish.
Then $P=0$.
\end{Lem}
\par\begin{proof}
We take $P$ as given by (\ref{TypicalElement}). Differentiation does not
change the total degree of a product $\sigma_j$, and hence only a fixed set
of such products appear in the derivatives of $P$. These derivatives
therefore give us a linear system from which the
trigonometric products may be eliminated.

So $P$ satisfies a differential equation of the form $\sum_0^M h_mP^{(m)}=h$
with $h,h_0\dts h_m\in\cH$. However $h$ must vanish on $\{z_n\}$, which
forces it to be zero.

Thus the equation for $P$ is linear and homogeneous. But the zero function
is obviously a solution and uniqueness forces this to be $P$.
\end{proof}

The results of Lemma~\ref{HFsmalls_lem} relating to differentiation
generally fail in $\cH_\cG$. The problem quite simply is that
$h\sin G$ may have $\gamma_0$ smaller than that of other terms
under consideration, but if $G$ is large its derivative may dominate the
other derivatives. Because the derivative of $G$ plays such an important
role what follows we introduce some special notation. 
\begin{Def}\label{flutter_def}
We write $g=G^\prime$ and call $\gamma_0(g)$ the {\em flutter} of $h\sin G$
and of $h\cos G$. The flutter of an element of $\cH$ is undefined.
\end{Def}
We do not make a general extension of the notion of flutter even to sums
because if $h_2=o(h_1)$ and $G_1=o(G_2)$ it is not clear what the flutter
of $h_1\sin G_1+h_2\sin G_2$ should be. However we do refer to the maximum
flutter of a sum, this being the maximum of the flutters of the summands.
\subsection{Algebraic independence}\label{algindep_subsec}
\begin{Lem}\label{injection_lem}
Let $G_\nu,\:\nu=1\dts u$ and $\lambda_{\mu,1}\dts\lambda_{\mu,n_\mu}$
be as in (\ref{STDefn}), and let $X_1\dts X_n$ be indeterminates.
Then the projection from $\cH[X_1\dts X_n]$ onto the ring
\[ \cH[\sin(\lambda_{1,1}G_1)\dts\sin(\lambda_{1,n_1}G_1)
\dts\sin(\lambda_{u,1}G_u)\dts\sin(\lambda_{u,n_u}G_u)] \]
is injective.
\end{Lem}

\par\begin{proof}
Write
\[
\Phi(X_1\dts X_n)=\sum_1^I f_iX_1^{\alpha_{i,1}}\cdots X_n^{\alpha_{i,n}},
\]
with the $f_i$ in $\cH$ and the $\alpha_{i,j}$ in $\N$, and suppose that
\[ \Phi(\sin(\lambda_{1,1}G_1)\dts\sin(\lambda_{1,n_1}G_1)
\dts\sin(\lambda_{u,1}G_u)\dts\sin(\lambda_{u,n_u}G_u))=0. \]
If the $f_i$ are not all zero, we may divide through the sum by an $f_i$
of maximal $\gamma_0$ to obtain a relation of the form
\begin{equation}\label{inj_lem_eq}
\sum_1^I c_j\sin^{\alpha_{j,1}}(\lambda_{1,1}G_1)\cdots\sin^{\alpha_{j,u}}(\lambda_{u,n_u}G_u))\lt 0,
\end{equation}
where the $c_j$ are non-zero real constants.

Theorem~2 of \cite{SaSh10} now shows that the left-hand side of (\ref{inj_lem_eq})
is a non-constant element of $\coef_{G_1^{-1}}$. The statement of the theorem does
not say that the element is non-constant, but the most cursory inspection of the
proof makes it clear that it is; alternatively we may appeal directly to
\cite{vdHoeven97ii}, which Theorem~2 of \cite{SaSh10} uses.

On taking $l=0$ in (\ref{coeffdefeq1}),
we get a contradiction. So the $f_i$ must all be zero, and we have established
Lemma~\ref{injection_lem}.
\end{proof}

Thus the elements of $\{\sin(\lambda_{1,1}G_1(x))\dts$ $\sin(\lambda_{u,n_u}G_u(x))\}$
are algebraically independent over $\cH$; c.f. \cite{Ax71}.

In the next subsection, we give a sort of asymptotic analogue of this.

\subsection{Asymptotic independence}\label{asymindsubsec}
We begin with a simple lemma.
\begin{Lem}\label{CoeffAsymLem}
Let $c_1$ and $c_2$ be non zero elements of $\R[\cS_\cT]$ and suppose that
$c_1-c_2\prec  c_1$. Then $c_1=c_2$.
\end{Lem}
\begin{proof}
From Theorem~2 of \cite{SaSh10}, or Theorem~7 of \cite{vdHoeven97ii}, $c_1-c_2$
is an element of $\coef_\cH$. Since it tends to zero it can only be the zero
element.
\end{proof}
\begin{Lem}\label{indep_prods_lem}
Suppose that $p_0\dts p_N\in\cH$ and let $\sigma_0\dts\sigma_N$ be pairwise
distinct products of elements of $S_\cT$ with no squares or higher powers
of cosines. Then
\begin{equation}\label{indep_prods_eq}
\gamma_0\left (\sum_{r=0}^Np_r\sigma_r\right )=\max\{\gamma_0(p_r);\:r=0\dts N\}.
\end{equation}
\end{Lem}
\par\begin{proof}
The point of the lemma is that there can be no partial cancellation
between different summands.
Clearly it is enough to prove the result for those $p_r$ of maximal
$\gamma_0$. Let $h$ be one such. Then we may assume each of the others
is a constant multiple of $h$, since if $l=\lim\{p_r/h\}$ the
terms involving $p_r-lh$ are $o(h)$ and will not affect the conclusion.
They may be therefore be removed from the sum on the left of
(\ref{indep_prods_eq}).

As in Lemma~\ref{injection_lem}, this sum must now be now equal to $h$
times a non-zero element of $\coef_\cH$. Thus the left-hand side of
(\ref{indep_prods_eq}) has $\gamma_0$ equal to $\gamma_0(h)$, as required.
\end{proof}

\begin{Th}\label{RatFnsTh}
Let $P$ be as in (\ref{TypicalElement}) and let $Q$ similarly be written
$Q=\sum_kq_k\tau_k$.
Suppose that  $\max_j\{\gamma(p_j)\}\neq\max_k\{\gamma(q_k)\}$ unless
both are equal to $\gamma(1)$. Suppose also that $P$ and $Q$ have no
common factor and that $P/Q$ is not of the form $K+o(1)$, $K\in\R\upstar$.
Then
\[ \gamma_0\left (\left ({P}/{Q}\right )^\prime\right )=\max_{j,k}
\{\gamma_0(p_j^\prime q_k),\gamma_0(p_jq_k^\prime),\gamma_0(p_jq_kg^P_j),
\gamma_0(p_jq_kg^Q_k)\}-2\max_k\{\gamma_0(q_k)\}, \]
where $g^P_j$ and $g^Q_k$ are of maximal flutter for the factors of $\sigma_j$
and $\tau_k$ respectively.
\end{Th}
Note that without the condition $\max_j\{\gamma(p_j)\}\neq\max_k\{\gamma(q_k)\}$
the result can fail even for exp-log functions, e.g. $P=x^2e^x+1$, $Q=xe^x+1$.
Of course if $\max_j\{\gamma(p_j)\}=\max_k\{\gamma(q_k)\}\neq \gamma(1)$ one
can divide through top and bottom by the appropriate power of a maximal $p_j$.
The example $P=x^2\sin x+x$, $Q=x^2\sin x+1$ shows the requirement that $P/Q$
is not of the form $K+o(1)$ is necessary.

\begin{proof}%[of Theorem~\ref{RatFnsTh}]
The assertion of the theorem amounts to saying that there
is no partial cancellation between the various terms $p_j^\prime q_k,p_jq_k^\prime,
p_jq_kg^P_j,p_jq_kg^Q_k$  of maximal $\gamma_0$ which arise in
$P^\prime Q-PQ^\prime$. We use the word `cancel' in this sense throughout
the proof.

The thrust of the argument is that if there is cancellation then we can
find simpler $P$ and $Q$ where there is cancellation. As in
Lemma~\ref{indep_prods_lem}, we shall remove terms of less than maximal
$\gamma_0$ and divide out common factors. It would be tedious to change the
notation every time we do this, so we shall generally suppose that the removed
terms were never present in the first place.

We have
\begin{equation}\label{num_expanded}
\hspace{-3mm} P^\prime Q-PQ^\prime=\left (\sum_jp_j^\prime\sigma_j+\sum_jp_j\sigma_j^\prime
\right )\sum_kq_k\tau_k-\sum_jp_j\sigma_j\left (\sum_kq_k^\prime\tau_k
+\sum_kq_k\tau_k^\prime\right ).
\end{equation}
Lemma~\ref{indep_prods_lem} assures us that cancellation can only take place
between terms where the trigonometric products are the same. In particular the
sets of arguments to trigonometric functions in the products $\sigma_j\tau_k$
must be identical, and this set is unchanged by differentiation of $\sigma_j$
or $\tau_k$. So we may assume that the same argument sets occur in the
terms of (\ref{num_expanded}).

We first consider the case when  neither $\max\{\gamma_0(p_j)\}$ nor
$\max\{\gamma_0(q_k)\}$ is zero.
Then we claim that only $p_j$ of maximal $\gamma_0$ can contribute
to (\ref{num_expanded}). For with regard to $\sum_jp_j^\prime\sigma_j$,
Lemma~\ref{HFsmalls_lem}(ii) implies that non-maximal $p_j$  will have
non-maximal derivatives. In $\sum_jp_j\sigma_j^\prime$ any $g_l^P$ which appears
in a product of the sum $\sigma_j^\prime$ attached to a non-maximal $p_j$
will similarly appear attached to a maximal $p_j$. The same remarks apply to
the $\tau_k$.

Thus by ignoring terms of non-maximal $\gamma_0$, we may suppose that
$p_j\asymp p_m$, for each $j$, where $p_m$ is of maximal $\gamma_0$.
If $\omega_j$ is the
limit of $p_j/p_m$, then $p_j-\omega_jp_m$ may be ignored since it
is $o(p_m)$. The same argument applies to the $q_k$ and hence we may take
$p_j=\omega_jp_m$ for all $j$, and $q_k=\mu_kq_m$ for all $k$, with
the $\omega_j,\mu_k\in\R\upstar$. Then
\begin{eqnarray}\label{P-Q_dash-eq}
P^\prime Q-PQ^\prime &\asymp &\{p_m^\prime q_m-p_mq_m^\prime\}\sum_j\omega_j\sigma_j
\sum_k\mu_k\tau_k\nonumber \\
&+&p_mq_m\left\{\sum_j\omega_j\sigma_j^\prime
\sum_k\mu_k\tau_k-\sum_j\omega_j\sigma_j\sum_k\mu_k\tau_k^\prime\right \}.
\end{eqnarray}
We show that this relation is also valid in the other cases. For suppose
that $\max\gamma_0(p_j)=0\neq\max\gamma_0(q_k)$. Since every $G_\nu$ has
strictly positive $\gamma_0$, Lemma~\ref{HFsmalls_lem}(ii) implies that
every $\gamma_0(p_j^\prime)$ is less than the flutter of each element of
$S_\cT$. Hence the entire sum
$\sum_jp_j^\prime\sigma_j$ can be removed from (\ref{num_expanded}) since it is
$o(p_j\sigma_j^\prime)$ for each $j$. The argument of the previous case now
applies to the remaining terms, so we may take $p_j=\omega_jp_m$ and
$q_k=\mu_kq_m$, with $\omega_j,\mu_k\in\R\upstar$. We obtain (\ref{P-Q_dash-eq})
without the leading $p_m^\prime q_m$, but of course it could be added since
the terms of maximal $\gamma_0$ are unaffected.

If $\max\{\gamma_0(q_k)\}=0\neq\max\{\gamma_0(p_j)\}$ or if
$\max\{\gamma_0(p_j)\}=0=\max\{\gamma_0(q_k)\}$, the argument is very similar.
So (\ref{P-Q_dash-eq}) holds in all cases.

Now let $s$, respectively $c$, be a sine or cosine of non-maximal flutter
occurring in $P$. Any $s^\prime$ or $c^\prime$ appearing in (\ref{P-Q_dash-eq})
will give a term of non-maximal $\gamma_0$, which may be removed.
Then if we rewrite (\ref{P-Q_dash-eq}) grouping together terms with the same
trigonometric power product, Lemma~\ref{indep_prods_lem} shows that any
cancellation must be within these groups. The presence of powers of $s$,
perhaps with a $c$, has no effect on such cancellation, and it follows
that $s$ and $c$ may be removed from (\ref{P-Q_dash-eq}).
Hence we may assume that all the sines and cosines in (\ref{P-Q_dash-eq})
have maximal flutter.

At this point the argument breaks into nine cases according to the order
relations between $\gamma(p_m)$ and $\gamma(q_m)$, and between $\gamma_0(g_m)$
and $\gamma_0(g_M)$. The interchangeability of $P$ and $Q$ reduces that
to five with varying degrees of similarity. We give a detailed proof when
$\gamma(p_m)>\gamma(q_m)$ and $g_m\succ g_M$ and then indicate
more briefly what happens in the other cases.

The inequality $\gamma(p_m)>\gamma(q_m)$ implies $p_m^\prime q_m\succ  p_mq_m^\prime$
by Lemma~\ref{HFsmalls_lem}(iii). So we may remove $p_mq_m^\prime$ from the right
of (\ref{P-Q_dash-eq}). Likewise we may remove the sum $\sum_k\mu_k\tau_k^\prime$
since $g_m\succ g_M$. This gives
\begin{equation}\label{RatFnsIAeq}
P^\prime Q-PQ^\prime \asymp q_m\sum_k\mu_k\tau_k\left \{p_m^\prime\sum_j\omega_j
\sigma_j+p_m\sum_j\omega_j\sigma_j^\prime\right \}.
\end{equation}
Lemma~\ref{indep_prods_lem} shows that if there is cancellation in
(\ref{RatFnsIAeq}) it cannot be in $\sum_k\mu_k\tau_k$. Hence it would have
to be in $p_m^\prime\sum_j\omega_j\sigma_j+p_m\sum_j\omega_j\sigma_j^\prime$.
It follows that $p_m^\prime\sim p_mg_m$ and
\[ \sum_j\omega_j\sigma_j+g_m^{-1}\sum_j\omega_j\sigma_j^\prime\prec 
\sum_j\omega_j\sigma_j. \]
$\sum_j\omega_j\sigma_j$ and  $g_m^{-1}\sum_j\omega_j\sigma_j^\prime$ are elements
of $\R[\cS_\cT]$ and so Lemma~\ref{CoeffAsymLem} implies that
$\sum_j\omega_j\sigma_j= -g_m^{-1}\sum_j\omega_j\sigma_j^\prime$.
Now the quotient of these sums can
be integrated on intervals where $\sum_j\omega_j\sigma_j$ has no zeros, to give
$\sum_j\omega_j\sigma_j=Ke^{-G_m}$, $K\in\R\upstar$. By Lemma~\ref{zerolts_lem},
these intervals abut, each one to the next. Continuity at the end points would
show that $K$ is independent of the interval of integration, but the whole
thing is completely impossible anyway. This proves Theorem~\ref{RatFnsTh}
for this case.

Now suppose that $\gamma(p_m)<\gamma(q_m)$ and $g_m\succ g_M$.
This time we remove $p_m^\prime q_m$ and $\sum_k\mu_k\tau_k^\prime$ from the right
of (\ref{P-Q_dash-eq}), and factor out $p_m\sum_k\mu_k\tau_k$. We see that any
cancellation must occur in
$q_m(\sum_j\omega_j\sigma_j)^\prime -q_m^\prime\sum_j\omega_j\sigma_j$.
As in the previous case, this leads to $q_m^\prime\sim q_mg_m$ and 
$(\sum\omega_j\sigma_j)^\prime-g_m\sum\omega_j\sigma_j\prec g_m\sum\omega_j\sigma_j$.
A contradiction is obtained as before.

When $\gamma(p_m)<\gamma(q_m)$ and $g_m\asymp g_M$, we may take $g_M=g_m$.
Then similar reasoning leads to $\Sigma_1\Sigma_2^\prime =g_m\Sigma_1\Sigma_2+
\Sigma_1^\prime\Sigma_2$,
with $\Sigma_1=\sum_j\omega_j\sigma_j$, $\Sigma_2=\sum_k\mu_k\tau_k$.
We divide through and integrate on intervals where $\Sigma_1\Sigma_2$ has no zeros.
This time we obtain $\sum_k\mu_k\tau_k=Ke^{G_m}\sum_j\omega_j\sigma_j$ with
$K\in\R\upstar$, and again the theorem follows in this case.

The remaining possibility is that $\gamma(p_m)=\gamma(q_m)=\gamma(1)$.
Now we have (\ref{P-Q_dash-eq}) with $p_m,q_m\in\R$. The first summand of
(\ref{P-Q_dash-eq}) disappears and the familiar arguments give
$\Sigma_1^\prime\Sigma_2=\Sigma_1\Sigma_2^\prime$. Integration as before shows that
$\Sigma_1=K\Sigma_2$, $K\in\R\upstar$. On taking account of removed terms,
we see that cancellation would imply $P=Q(K+o(1))$, which was prohibited.
\end{proof}

\section{Asymptotic domains}\label{AsymFldsSec}
The author's basic approach to generating an expansion of a function has
been to build a tower of fields
\begin{equation}\label{fieldtower}
\R=\mathcal{F}_0\subset\R(x)=\mathcal{F}_1\subseteq\mathcal{F}_2
\subseteq\cdots\subseteq\mathcal{F}_N,
\end{equation}
with each $\mathcal{F}_j$ a simple extension $\mathcal{F}_{j-1}(f_i)$ of its
predecessor and the given function an element of $\mathcal{F}_N$. The types of
function $f_j$ permitted determines the class of functions that can
be expanded. In the present context, this was first done for {\em exp-log}
functions in \cite{Richardsonetal}, with each $f_j$ allowed to be an exponential
or a logarithm of an existing element of $\mathcal{F}_{j-1}$. 

One problem that may arise in this area is that of `indefinite cancellation'.
If one expands sub-expressions without due care, the top-level terms may
cancel indefinitely higher in the expression tree.
See \cite{Sh1,Richardsonetal,Gruntz96,Sh_bk}, and also the example in
Section~\ref{expn_comments}. For exp-log functions the problem
may be overcome as follows.

Suppose that we have a scale $\{\bt_1\dts\bt_m\}$ as in
Definition~\ref{scaledef}, and an exp-log function $f=f(\bt_1\dts\bt_m)$.
We divide $f$ by a $\{\bt_1\dts\bt_m\}$-monomial to make $f\asymp 1$.
The first coefficient in the $\bt_m$ expansion of $f$ may then be obtained as
$f(\bt_1\dts\bt_{m-1},0)$; for the second we differentiate
with respect to the last argument before replacing $\bt_m$ by zero, and so on.
Expansions with respect to $\bt_1\dts\bt_{m-1}$ are handled similarly, etc.

An important point is that we have closed forms for the coefficients in this
situation and these and their algebraic combinations can be tested for zero
equivalence.

When integration is added into the signature, the device of replacing a scale element
by zero is problematic to say the least. Thus given that $\int x^{-1}=\log x$ for
example,  we cannot sensibly replace $\bt_m$ by zero if $\bt_m=x^{-1}$. For these
and other reasons, the concept of an asymptotic field was introduced in
\cite{Sh7}; see also \cite{Sh_bk}, \S 5.2.

The idea is to equip each field, $\mathcal{F}_j$ of (\ref{fieldtower}) with
a collection of subfields called shadow fields, one for each scale
element. Then instead of replacing a scale element by zero, we project onto
the appropriate shadow field to obtain the coefficient. Shadow fields are
defined in such a way that problems of indefinite cancellation do not
thereby arise.  Of course when the addition of an element gives another
asymptotic field the process may be repeated with a new element.

Our present aim is to extend these ideas to the measured case. Because we are
unable to handle integrals of quotients of elements of $\cH_\cG$ in
any generality and we only need to divide by elements of $\cH$ in what follows,
it seems best here to work with domains rather than fields. Thus we need to
handle the addition of integrals of the forms $\int h\sin G$,
$\int h\cos G$, with $h,G\in\cH$, to an existing asymptotic domain $\cA=\cH_\cG$;
see Definition~\ref{Ch5-AsymDomainDef} below. This will allow measured multiseries
to be computed for rational combinations of such integrals over that domain.

\subsection{Shadow domains}
\begin{Def}\label{RIdef}
Let $\bt$ be an element of $\cH$ which tends to zero. We write
$\cR_\bt(\cA)=\{a\in\cA; \forall\varepsilon\in\R^+, a\prec \bt^{-\varepsilon}\}$
and
$\cI_\bt(\cA)=\{a\in\cA:\exists\delta\in\R^+, a\prec \bt^\delta\}.$
\end{Def}
Usually $\bt$ will be a scale element.
These definitions parallel those in the Hardy-field case, but $\cR_\bt(\cA)$
and $\cI_\bt(\cA)$ exhibit slight differences in the new setting.
\begin{Lem}\label{RILem}
$\cR_\bt(\cA)$ is a subring of $\cA$ and $\cI_\bt(\cA)$ is a prime ideal of
$\cR_\bt(\cA)$.
\end{Lem}
The proofs follow those of the Hardy-field case and are easy anyway.
However we do not assert that $\cR_\bt(\cA)$ is a differential subring
of $\cA$ nor that $\cI_\bt(\cA)$ is a differential ideal of $\cR_\bt(\cA)$,
for the flutter might intervene. For example $\sin e^x$ belongs to
$\cR_{1/x}$ but its derivative does not.

\begin{Def}\label{ShadowDef}
Let $\cA$ and $\bt$ be as above, and let $\cS$ be a subdomain of
$\cA$ containing $\R$. We say that $\cS$ is a
{\em shadow domain} with respect to $\bt$ if the following hold:\\
{\rm SF(i)}. $\cS\cap\cH$ is closed under relative differentiation; that is to say
if $a$ and $b$ belong to $\cS\cap\cH$ and $b^\prime\neq 0$ then
$a^\prime\in b^\prime\cS\cap\cH$.\\
{\rm SF(ii)}. $\cS\cap\cI_\bt(\cA)=\{0\}$.
\end{Def}
When $\gamma(\bt)>\gamma(x)$, SF(i) is equivalent to closure of $\cS\cap\cH$
under differentiation, with the reasonable assumption that $x\in\cS$ in this
case. However $\log x$ may well belong to $\cS\cap\cH$ with
$\bt=x^{-1}$. For $\cS$ itself, even closure under relative differentiation
might be prohibited by the flutter, but SF(i) as given turns out to be sufficient
for our purposes.

SF(ii) ensures that coefficients in an expansion which lie in the same shadow
subdomain cannot partially cancel to leave a small residue. Such an eventuality
could lead to indefinite cancellation.

It is immediately clear that the union of an increasing chain of shadow domains
with respect to $\bt$ has itself the shadow property. So it makes sense to
consider maximal shadow domains with respect to $\bt$.
\subsection{Shadow expansions}
There are two more definitions to make before we can begin to consider adding
integrals to our algebra $\cA$. The following is an adaption of the
Hardy-field version from \cite{Sh7} and \cite{Sh_bk}.
\begin{Def}     \label{Ch5-AsymDomainDef}
Let $\cA$ be as above. We say that $\cA$
is an {\em asymptotic domain} if the following conditions hold:
\begin{enumerate}
\item $\cA$ contains a scale $\{\bt_1,\ldots,\bt_m\}$ satisfying
Definition~\ref{scaledef}.
\item There are subdomains ${\mathcal S}_1\subset{\mathcal S}_2
\subset\cdots\subset{\mathcal S}_m$ of $\cA$ such that each ${\cal S}_i$
is a shadow domain with respect to $\bt_i$.

Also $\bt_j\in{\mathcal S}_i$ whenever $j<i$
and moreover $\gamma(1),\gamma(\bt_1)\dts\gamma(\bt_{i-1})$ are the
comparability classes of ${\mathcal S}_i\cap\cH$.
\item For any given element of $\cA$, we can compute a measured
multiseries expansion with each term a product of an element of the
shadow domain $\cS_i$, for the appropriate $i$, and powers of the
scale elements $\bt_i\dts\bt_m$.
\end{enumerate}
\end{Def}
A computable multiseries expansion whose coefficients lie in
shadow domains is called a {\em shadow expansion}. Projecting a
suitable function $f$ onto a shadow domain ${\cS}_i$ is equivalent
to giving the first term of the $\bt_i$-expansion of $f$. 

The following parallels Lemma~17 of \cite{Sh_bk}.
\begin{Lem}\label{asymdomcond_lem}
Let $\cA$ be an asymptotic domain. Then for each $i=1\dts m$, there
is a computable homomorphism $\eta_i$ from $\cR_i$ to $\cS_i$ such that for
each $f\in\cR_i$, we have $f-\eta_i(f)\in \cI_i$.

Moreover for $1\leq j<i\leq m$, there is a computable homomorphism $\eta_{j,i}$
from $\cR_j(\cS_i)$ to $\cS_j$ such that $\eta_{j,i}\circ\eta_i=\eta_j$. In
addition if $1\leq k<j<i\leq m$, then $\eta_{k,j}\circ\eta_{j,i}=\eta_{k,i}$.

Conversely let $\cA$ be a subdomain of $\cX$ with $\cA$ an algebra over
$\cH$ satisfying (i) and
(ii) of Definition~\ref{Ch5-AsymDomainDef}. Suppose that there are computable
functions $\eta_i$, $i=1\dts m$ and $\eta_{j,i}$, $1\leq j<i\leq m$ satisfying
the conditions of the above paragraph. Suppose also that for any given $f\in\cA$,
we can compute a $\bt_1\dts\bt_m$-monomial, $\tau$, and an element, $c$, of the set
$\cA\cap\coef_\cH$ such that $f\sim c\tau$. Then $\cA$ is an asymptotic domain.
\end{Lem}
\par\begin{proof}%[Lemma~\ref{asymdomcond_lem}]
Basically if $f\in\cR_i$ and the leading term of its multiseries contains
positive powers of any of $\bt_i\dts\bt_m$ then $\eta_i(f)=0$. Otherwise
$\eta_i(f)$ is the leading term in the $\bt_i$ expansion. The details are
very similar to those in \cite{Sh_bk} and are not repeated here.
\end{proof}

Again as in \cite{Sh7} and \cite{Sh_bk}, we call $\eta_i(f)$ the $i$-th
{\em shadow} of $f$ and $\xi_i(f)=f-\eta_i(f)$ the $i$-th {\em ghost}.
When $f\asymp T$ with $T$ a $\bt_i\dts\bt_m$ monomial, we shall often
write $f_i$ for $T\eta_i(T^{-1}f)$, and with a slight abuse of notation
refer to $f_i$ as the shadow of $f$ even when $f$ is not in $\cR_i$.

If take each $\cS_i$ to be maximal, then $\cS_i$ will be the subdomain of
$\cA$ consisting of those elements having $\{\bt_1,\ldots,\bt_{i-1}\}$
measured multiseries. We shall assume henceforth that this is the case.

Suppose that $\cS_i$ is a shadow domain of $\cA$ with respect to
$\bt_i$ and that some $F\in\cA$ has a $\{\bt_1,\ldots,\bt_{i-1}\}$ shadow
expansion. Since multiseries expansions may be combined algebraically as
in \cite{Sh_bk}, \S 4.3, it follows that $\cS_i(F)$ has the shadow
property with respect to $\bt_i$.

Now suppose we have an asymptotic domain $\cA$ and we wish to add a new
element $f$. The idea is as follows.
\begin{enumerate}
\item We compute a power product of scale elements $T$ such that
$T^{-1}f\asymp 1$, adding a new scale element if necessary.
\item For each scale element $\bt_i$, we define a shadow $\eta_i(T^{-1}f)$
and prove that $\cS_i(\eta_i(T^{-1}f))$ is a shadow domain.
\item We give a method for computing $\bt_j$-shadows of $\bt_i$-shadows,
$j<i$, and $\bt_j$-shadows of $\bt_i$-ghosts, $j\leq i$. We also show that
$\eta_{j,i}\circ\eta_i=\eta_j$ for $j<i$ and that
$\eta_{k,j}\circ\eta_{j,i}=\eta_{k,i}$ for $k<j<i$.
\item To expand $F\in\cA(f)$ we first divide $F$ by the leading monomial of
its multiseries. The beginning of the $\bt_m$ expansion is obtained as the
$\bt_m$-shadow $F_m$ plus $\bar{T}\eta_m(\bar{T}^{-1}\xi_m(F))$, where
$\bar{T}$ is the product of scale elements having the same $\gamma_0$ as
$\xi_m(F)$, and we continue in this way.
\item The $\bt_{m-1}$ expansion of a coefficient in the $\bt_m$ expansion
is obtained in an entirely analagous fashion, and similarly for the other
$\bt_i$.
\end{enumerate}
It will turn out that we only need to add a new scale element when
an element of $\cH$ is integrated to give a new iterated logarithm.

It is not hard to see that the rational functions of $x$ form an
asymptotic field, and \cite{Sh7} showed how to add exponentials, integrals
and algebraic roots in the Hardy-field case. \cite{Sh9} allowed a variety
of other functions to be added including trigonometric functions whose
arguments tend to a finite limit. Those whose arguments tend to infinity
can be handled as in \cite{SaSh10}. So $\cA$ is an
asymptotic domain provided that $\cH$ is an asymptotic field.

It is perhaps worth stressing the point that when all multiseries terms of
$G$  tend to infinity, we have to take the shadows of $\sin G$ and
$\cos G$ to be the functions themselves.
For example
\begin{equation}\label{ghostgexample}
\frac{\sin(x+\log x)}{\log x}+\frac{\cos(x+\log x)}{\log^2x}+\cdots 
\end{equation}
is a  perfectly valid beginning of a measured multiseries and we cannot
sensibly change anything when we take the $x^{-1}$-shadow. However
when we differentiate, the trigonometric functions introduce a factor
$1+x^{-1}$ and clearly the $x^{-1}$ cannot now be part of the
$\log^{-1}x$-shadow. We look again at this problem at the start of
Section~\ref{AsymFldInt}.

Our present concern is the addition of $\int h\sin G$ and $\int h\cos G$.
The difficult step is (ii) above.
\section{Adding integrals}\label{AddIntsSec}
If we want to add an integral, the first thing we need to compute is a
monomial asymptotic to it. For the Hardy-field case, this is
provided by the results of Section~IV in \cite{Hardy12}. We shall
follow the convention there that $\int f$ means $\int^xf$ if this
tends to infinity and $\int_x^\infty f$ otherwise. As a result
we never have $\int f\asymp 1$.

We shall need the following, which is essentially from \cite{Hardy12}.
\begin{Lem}\label{IntEqLem}
Let $\phi,\psi$ be elements of a Hardy field which contains $\int\phi$,
$\int\psi$ and $\int \phi\psi$.
\begin{enumerate}
\item If $\phi\asymp\psi$ then $\int\phi\asymp \int\psi$; if $\phi\prec \psi$
then $\int\phi\prec \int\psi$.
\item If $\gamma(\psi)<\gamma(\int \phi)$ then $\int \phi\psi\asymp \psi\int \phi$.
\item If $\gamma(\psi)=\gamma(\int \phi)$ then $\int \phi\psi\preceq \psi\tau\int \phi$
for some $\tau$ with $\gamma(\tau)<\gamma(\psi)$.
\item If $\gamma(\psi)>\gamma(\int \phi)$ then 
$\int \phi\psi=o(\psi\int \phi)$.
\item In all cases $\int \phi\psi\preceq \psi\tau\int \phi$ with $\gamma(\tau)<\gamma(\psi)$.
\end{enumerate}
\end{Lem}
\par\begin{proof}%[of Lemma~\ref{IntEqLem}]
We may take $\phi$ and $\psi$ to be positive. Then (i) and (ii) follow easily from
Lemma~\ref{HFsmalls_lem}.

If $\gamma(\psi)=\gamma(\int \phi)$, let $\int \phi=\psi^r\chi$ with
$r\in\R\upstar$, and $\gamma(\chi)<\gamma(\psi)$. Then if $r\neq -1$ we have
$\phi\asymp  \psi^{r-1}\psi^\prime\chi$, and hence
\[ \int \phi\psi\asymp  \int\psi^r\psi^\prime\chi\asymp\psi^{r+1}\chi\asymp\psi\int \phi, \]
by part (ii).

If $r=-1$ take the case $\psi\lt 0$ and suppose that
$\int\phi\psi\succ\psi^{1-2\delta}\int\phi$ with $\delta\in\R^+$. Then
$\int\phi\psi\succ\psi^{-\delta}$, and differentiation of this gives
$\phi\succ\psi^\prime\psi^{-2-\delta}$. However
$\phi=(\int\phi)^\prime=(\psi^{-1}\tau)^\prime\asymp\psi^\prime\psi^{-2}\tau$, which
yields the contradiction $\tau\succ\psi^{-\delta}$; whence the conclusion.
The case when $\psi\lt\infty$ is similar.

Now suppose that $\gamma(\psi)>\gamma(\int \phi)$. Then
Lemma~\ref{HFsmalls_lem}(iii) implies
$\phi\psi\prec \psi^\prime\int \phi=(\psi\int \phi)^\prime -\phi\psi$ and this gives
$\int \phi\psi\prec \psi\int \phi$; thus we have established (iv), and
(v) is now immediate.
\end{proof}

In our present situation, account needs to be taken of the flutter of coefficients.
We shall concentrate on $\int h\sin G$, where $h,G\in\cH$ and $G$ has all
its multiseries terms tending to infinity, as at the beginning of
Section~\ref{AsymValSec}. It is clear that $\int h\cos G$ may be treated
in exactly the same way as $\int h\sin G$, and we shall sometimes make use
of this. We recall that $g=G^\prime$ and for convenience take $h$ and $g$ to be
positive. We simplify notation by writing $H=\int h$, $S=\sin G$ and
$C=\cos G$.

If $h=g(K+\varepsilon)$, $K\in\R\upstar$, $\varepsilon\lt 0$, we use
the closed form $\int g\sin G=-\cos G$ and proceed with
$\int g\varepsilon S$. So we assume that $h\not\asymp g$.

The basic method of getting an $\cH$-monomial asymptotic to $\int hS$ is to
integrate by parts. Suppose first that $(h/g)^\prime h^{-1}\lt 0$; we call
this the {\em Diff-h}~case. Then
\begin{equation}\label{Diff-h}
\int hS=\int\frac{h}{g}gS=-\frac{hC}{g}+\int\left (\frac{h}{g}\right )^\prime C.
\end{equation}
Before we can deduce that $\int hS\asymp  -g^{-1}hC$ we need to show that
if $\tilde{h}=o(h)$ then $\int \tilde{h}C=o(\int hS)$.
This forms part of the corollary to Theorem~\ref{int_flutter_th} below.

If we take $h_{\langle 0\rangle}=h$ and $h_{\langle j+1\rangle}=(h_{\langle j\rangle}/g)^\prime$
for $j\geq 0$, we can continue to integrate by parts in the same sense while
$h_{\langle j+1\rangle}=o(h_{\langle j\rangle})$.
This will often be the case, since provided that the conditions of
Lemma~\ref{HFsmalls_lem}(viii) continue to be met, we have
\begin{equation}\label{hseqcondition}
\frac{h_{\langle j+2\rangle}}{h_{\langle j+1\rangle}}=
\frac{(h_{\langle j+1\rangle}g^{-1})^\prime}{(h_{\langle j\rangle}g^{-1})^\prime}
\bowtie\frac{h_{\langle j+1\rangle}g^{-1}}{h_{\langle j\rangle}g^{-1}}=
\frac{h_{\langle j+1\rangle}}{h_{\langle j\rangle}},
\end{equation}
for $j\geq 0$. In that case we can obtain a development of the form
\begin{equation}\label{h-seq}
\int hS=\sum_{n=0}^N (-1)^{n+1}\left (\frac{h_{\langle 2n\rangle}C}{g}-
\frac{h_{\langle 2n+1\rangle}S}{g}\right )+(-1)^{N+1}\int h_{\langle 2N+2\rangle} C.
\end{equation}
However (\ref{hseqcondition}) can fail. Let $h=x^3(1+e^{-x^3})$ and
$G=x^2/2$. Then $h_{\langle 1\rangle}=2x(1+e^{-x^3})-3x^4e^{-x^3}$, 
$h_{\langle 2\rangle}=e^{-x^3}(9x^5-15x^2)$ and
$h_{\langle 3\rangle}=e^{-x^3}(-27x^6+81x^3-15)$. So in this example we have
$h\succ  h_{\langle 1\rangle}\succ  h_{\langle 2\rangle}$,
but $h_{\langle 2\rangle}\prec  h_{\langle 3\rangle}$. When something like this
occurs, we can if necessary integrate by parts in the opposite sense.
This does not cause us to retrace our steps, since a constant will
effectively have been differentiated out of the lead term and integration
will not put it back because of our convention with integrals. Thus in the
example, we differentiated a 2 in passing from $h_{\langle 1\rangle}$ to
$h_{\langle 2\rangle}$. When processing $\int h_{\langle 1\rangle}C$ above
we have the exact form for $\int 2xC$, but we still need to continue
with integrating the other terms by parts.

Integration by parts in the reverse sense occurs when $h^{-1}Hg\lt 0$. 
We refer to this case as {\em Int-h}. We then obtain
\begin{equation}\label{Int-h}
\int hS=HS-\int HgC.
\end{equation}
We now write $H_{\langle 0\rangle}=H$ and $H_{\langle j+1\rangle}=\int H_{\langle j\rangle}g$,
$j=0\dts N$. Corresponding to (\ref{h-seq}), we have
\begin{equation}\label{IntH_seq}
\int hS=\sum_{n=0}^N (-1)^n\{H_{\langle 2n\rangle}S-H_{\langle 2n+1\rangle}C\}+
(-1)^{N+1}\int H_{\langle 2N+2\rangle}g S.
\end{equation}
Then for $j\geq 0$,
\[ \frac{H_{\langle j+2\rangle}}{H_{\langle j+1\rangle}}=
\frac{\int H_{\langle j+1\rangle}g}{\int H_{\langle j\rangle}g}\bowtie
\frac{H_{\langle j+1\rangle}}{H_j}. \]
For this time the conditions for Lemma~\ref{HFsmalls_lem}(viii) are guaranteed
by our convention regarding integrals, and so we remain in Int-h
for successive values of $j$.
It will follow from Theorem~\ref{int_flutter_th} that this is an
asymptotic expansion of $\int hS$ in the Int-h case.

\begin{Lem}\label{diffintcc_lem}
If, as we are assuming, $h\not\asymp  g$ then
\[ h^{-1}(h/g)^\prime\bowtie(h^{-1}Hg)^{-1}. \]
Also $h^{-1}Hg\lt K\in\R\upstar$ implies
$h^{-1}(h/g)^\prime\lt K^{-1}$.
\end{Lem}
\par\begin{proof}%[of Lemma~\ref{diffintcc_lem}]
Let $\lambda=h/g$. We have $h^{-1}Hg=H/\lambda$ and
$h^{-1}(h/g)^\prime=\lambda^\prime/H^\prime$. If $H\asymp  \lambda$ then
$h\asymp  (h/g)^\prime$ and so both $h^{-1}(h/g)^\prime$ and
$h^{-1}Hg=H\lambda^{-1}$ are asymptotic to non zero constants. Otherwise
Lemma~\ref{HFsmalls_lem}(viii) gives the first assertion. The second
follows from L'H\^{o}pital's Rule.
\end{proof}

Thus the Diff-h and Int-h cases are mutually exclusive and together
almost exhaust the possibilities.
It is not generally true that $h^{-1}(h/g)^\prime\asymp   ( h^{-1}Hg)^{-1}$.
For example if $H=\log_2x$ and $g=\log x$ then $h^{-1}(h/g)^\prime\asymp  
x^{-1}\log^{-1}x$ and $(h^{-1}Hg)^{-1}\asymp   x^{-1}\log^{-2}x\log_2^{-1}x$.

It remains to give a putative expansion when $(h/g)^\prime h^{-1}\lt
K\in\R\upstar$.
Let $\omega=(h/g)^\prime h^{-1}-K$. Two applications of integration
by parts in the Diff-h sense give
\begin{equation}\label{Kdiffeq}\hspace{-2mm}
\int hS=\frac{-hC}{(1+K^2)g}+\left (\frac{h}{g}\right )^\prime\frac{S}{(1+K^2)g}
+\frac{1}{1+K^2}\int\omega(2K+\omega) hS+\int\frac{\omega^\prime h}{g}S.
\end{equation}
Since $\omega\lt 0$ and $G\lt\infty$, we have $\omega^\prime=o(g)$. The
corollary to Theorem~2 will show that both the integrals on the right of
(\ref{Kdiffeq}) are $o(\int hS)$. It is possible that this particular case
may reoccur with one of the integrals on the right of (\ref{Kdiffeq}), and
we then repeat the process. Otherwise we can continue to integrate by parts
in whichever sense is applicable.

What we mostly want from (\ref{Diff-h}),
(\ref{Int-h}) and (\ref{Kdiffeq}) is to obtain an element of $\cH$ with
the same $\gamma_0$ as $\int hS$. The similarities between (\ref{Kdiffeq})
and (\ref{Diff-h}) allow us to subsume the present case into Diff-h
in what follows.
\subsection{A differential equation associated with $\int hS$}\label{desubsec}
We have no reason to believe that the series of (\ref{h-seq}) and (\ref{IntH_seq})
converge. However the expansions do suggest that we seek elements, $f_1, f_2$,
of a Hardy-field extension of $\cH$ such that
\begin{equation}\label{f1f2-eq}
\int hS = f_1S + f_2C+A,\quad A\in\R.
\end{equation}
If we differentiate (\ref{f1f2-eq}),
we get $hS=(f_1^\prime -f_2g)S+(f_1g+f_2^\prime)C$.
 In this situation, we may equate coefficients of $S$ and $C$, for if
$f_1g+f_2^\prime\neq 0$ we would obtain $\cot G$ equal to a Hardy-field
element. Hence
\begin{equation}\label{f1f2eqs}
h=f_1^\prime -f_2g\qquad \mbox{and}\qquad f_2^\prime=-f_1g.
\end{equation}
It now follows easily that any such $f_2$ must satisfy
\begin{equation}\label{f2diff-eq}
y^{\prime\prime}-y^\prime g^\Delta +yg^2+hg=0.
\end{equation}
Before grappling with this equation in its full generality, we
consider the case when $G(x)=x$.
Then (\ref{f2diff-eq}) becomes $y^{\prime\prime}+y+h=0$, and Theorem~3.9
of \cite{Bosh4} tells us that this has a solution in Hardy-field extension
of $\cH$. In some cases there is only one such solution, and we take $f_2$
to be that. In other cases all solutions lie in such extensions, and we
take $f_2$ to be any solution. Obviously two different solutions of
$y^{\prime\prime}+y+h=0$ differ by $AS+BC$ for some $A,B\in\R$, but nonetheless,
for certain $h$, each may lie in a (different) Hardy field; see \cite{Bosh4}.

Now for general $G$, substitute $G^{inv}(x)$ for $x$ and write
\[ \widetilde{\cH}=\cH\circ G^{inv}=_{def}\{f\circ G^{inv};\:f\in\cH\}. \]
Then $\widetilde{\cH}$ is closed under differentiation since $(G^{inv})^\prime=
(G^\prime\circ G^{inv})^{-1}$. Also composition with $G^{inv}$ does not affect the
ultimate sign of any $f\in\cH$. In other words $\widetilde{\cH}$ is a Hardy field.

We check that (\ref{f2diff-eq}) reduces to the form $y^{\prime\prime}+y+\phi=0$
under the substitution, with $\phi\in\widetilde{\cH}$. Let $F$ be any
solution of this differential equation lying in a Hardy--field extension of
$\widetilde{\cH}$ and let $Y=F\circ G$.
In general we are not entitled to conclude that $Y$ lies in a Hardy-field
extension of $\cH$, \cite{Bosh2} \S~11. We prove that here it does.

We may extend $\cH$ to a Hardy field containing $\sqrt{g}$, \cite{Robinson72}.
Our ability to calculate multiseries in $\cH$ will be unaffected, \cite{Sh7},
although we do not need to use this here.
The substitution $z=yg^{-1/2}$ transforms (\ref{f2diff-eq}) to
\[
z^{\prime\prime}+z\left (\frac{g^{\prime\prime}}{2g}-\frac{3(g^\prime)^2}{4g^2}
+g^2\right ) +h\sqrt{g}=0,
\]
and this satisfies the conditions of Theorem~3.1 in \cite{Bosh4}.
That theorem tells us that if $z$ is a solution which is consistent with
$\cH$ in the sense that elements of $\cH[z]$ have a definite sign for $x$
sufficiently large, then $z$ belongs to a Hardy-field extension of $\cH$.

However non-zero elements of $\widetilde{\cH}[F]$ do not have arbitrarily
large zeros, and so nor do non-zero elements of $\cH[Y]$. Thus $Yg^{-1/2}$
does indeed lie in a Hardy field extension of $\cH$. Hence so does $Y$,
and we may take $f_2=Y$.

If we put $f_1=-g^{-1}f_2^\prime$ and $f=f_1S+f_2C$, we see that
$f_1^\prime-f_2g=h$ and $f^\prime =hS$. So $f$ is an integral of $hS$,
and we have thus proved the following.
\begin{Th}\label{int_flutter_th}
Let $h$, $S$ and $C$ be as above.
Then there exist elements $f_1, f_2$ of a Hardy-field extension of
$\cH$ such that (\ref{f1f2-eq}) holds.
\end{Th}
We shall generally take $A=0$ in (\ref{f1f2-eq}) and regard $f_1S + f_2C$
as `the' integral of $hS$, just as one thinks of $\log x$ as being the
integral of $x^{-1}$. However we need to bear in mind that there is an
arbitrary constant.

In fact arbitrary constants pose a problem here just as they did in \cite{Sh7}.
Since the theory is about germs of functions at $+\infty$, we cannot expect to
specify a value of an integral at a finite point. So in general there may
be difficulty distinguishing different integrals, although in application
the matter might resolve.
We return to this question in the next section.

It is perhaps worth emphasising that when we add $\int hS$ to $\cA$, we are
only adding to $\cH$.

\begin{Cor} \label{Coroll1}
  Let $h$ be as above. Then $\int hC\asymp\int hS$.
\end{Cor}
\begin{proof}%[of Corollary 1]
  We check that $(f_1C-f_2S)^\prime=hC$. Then it follows immediately from
Lemma~\ref{indep_prods_lem} that $\int hC\asymp\int hS$.
\end{proof}
  
\begin{Cor} \label{Coroll2}
  Let $\widetilde{h}$ be another element of $\cH$ and take the arbitrary constant to
  be zero in integrals of the form (\ref{f1f2-eq}). If $\widetilde{h}\prec h$ then
$\int\widetilde{h}S=\prec\int hS$, and if $\widetilde{h}\asymp h$ then
$\int\widetilde{h}S\asymp \int hS$.
\end{Cor}
\begin{proof}%[of Corollary 2]
  We can extend $\cH$ again to contain $\widetilde{f}_1$ and $\widetilde{f}_2$
such that $\int\widetilde{h}S=\widetilde{f}_1S+\widetilde{f}_2C+\widetilde{A}$
with $\widetilde{A}\in\R$. Theorem~\ref{RatFnsTh} gives
\begin{equation} \label{Cor2Gam0h}
  \gamma_0(h)=\gamma_0((f_1S+f+2C)^\prime)=\max\{\gamma_0(\widetilde{f}_1^\prime),
  \gamma_0(\widetilde{f}_2^\prime),\gamma_0(\widetilde{f}_1g),\gamma_0(\widetilde{f}_2g)\}
\end{equation}
and similarly
\begin{equation} \label{Cor2Gam0hTilde}
  \gamma_0(\widetilde{h})=\max\{\gamma_0(f_1^\prime),\gamma_0(f_2^\prime),
  \gamma_0(f_1g),\gamma_0(f_2g)\}
\end{equation}
Suppose that $\widetilde{h}\prec h$ but $\int\widetilde{h}S\succeq\int hS$. So
$\max\{\gamma_0(f_1),\gamma_0(f_2)\}\leq\max\{\gamma_0(\widetilde(f)_1),\gamma_0(\widetilde(f)_2)\}$.

Suppose also that $\widetilde{f}_2\preceq \widetilde{f}_1$. Then the last two terms of the
maximum in (\ref{Cor2Gam0h}) are no greater than $\gamma_0(\widetilde{f}_1g)$.
If $\widetilde{f}_1\not\asymp 1$ then the first two are similarly no greater then
$\gamma_0(\widetilde{f}_1^\prime)$ by Lemma~\ref{HFsmalls_lem}ii.
But if $\widetilde{f}_1\asymp 1$ those first two are dominated by $\gamma_0(\widetilde{f}_1g)$
since $f_1$ and $f_2$ are derivatives of functions tending zero, while $g=G^\prime$ and $G\lt\infty$.
Hence $h\preceq\widetilde{h}$ contrary to hypothesis. Thus
$\widetilde{h}\prec h\implies\int\widetilde{h}S=\prec\int hS$.

If $\widetilde{h}\asymp  h$, we write  $\widehat{h}=Kh-\widetilde{h}$ with
$K=\lim(h^{-1}\widetilde{h})$ and use the previous case.
\end{proof}

We are now able to assert that the formulae (\ref{h-seq}), (\ref{IntH_seq})
and (\ref{Kdiffeq}) are genuine asymptotic expansions.

\section{The shadow domains of $\cA[\int hS]$}\label{AsymFldInt}
We take shadows with respect to $\bt_i$, as representing a typical scale
element. Recall that $\cS_i(\cA)$ is the shadow domain of $\cA$ with respect to
$\bt_i$, and $\eta_i$ is the projection from $\cA$ to $\cS_i(\cA)$.
Write $T_h$, $T_H$ and $T_g$ for the leading $\{\bt_1\dts\bt_m\}$
monomials in the multiseries of $h$, $H$ and $g$ respectively.
As before, we use $T$ to denote the monomial with the same $\gamma_0$
as $\int hS$; so in the Diff-h case, $T=T_h/T_g$.

Ghosts, $\xi_i(g)$, of $g$ can cause problems, as illustrated by (\ref{ghostgexample}).
We cannot replace $G$ in $S$ by its shadow $G_i$ without destroying the
basic periodicity, but $g$ often appears in expansions. The following
result provides a solution where it is needed.
\begin{Lem}\label{G_splitLem}
Let $G=G_1+G_2$ with $G_1\lt\infty$, $G_2\lt\pm\infty$, and write $g_1=G_1^\prime$
and $g_2=G_2^\prime$. Suppose that $g_2=o(g_1)$, and that
$\gamma(g_2/g_1)\geq \gamma(\bt_i)$. Then with $T$ as above,
\[\frac{1}{T}\int hS-\frac{\cos G_2}{T}\int h\sin G_1-
\frac{\sin G_2}{T}\int h\cos G_1\prec \bt_i^d \]
for some $d\in\R^+$, and similarly for $\int hC$.
\end{Lem}
\par\begin{proof}%[of Lemma~\ref{G_splitLem}]
We write $S_1=\sin G_1$, $S_2=\sin G_2$, $C_1=\cos G_1$ and $C_2=\cos G_2$.
On integrating by parts, we obtain
\begin{equation}\label{G-splitEq2}
\int hS_1C_2 = C_2\int hS_1+\int\left (g_2S_2\int hS_1\right ).
\end{equation}
We show that $T^{-1}$ times the second integral belongs to the $\bt_i$ ghost.
Parts applied to the inner integral gives
\begin{eqnarray} \hspace{-6mm}
\int \left (g_2S_2\int hS_1\right )&=&-\int \frac{g_2h}{g_1}C_1S_2+
\int\left (g_2S_2\int \left (\frac{h}{g_1}\right )^\prime C_1\right )\label{GSPL_insert}\\
&=&\frac{1}{2}\int\frac{g_2h}{g_1}\{\sin(G_1+G_2)-\sin(G_1-G_2)\}+
\int\left (g_2S_2\int \left (\frac{h}{g_1}\right )^\prime C_1\right )\label{G_splitEq3}
\nonumber
\end{eqnarray}
From here the argument splits into the Diff-h and Int-h cases. We begin by showing that
whichever of these holds for $\int hS$ also holds for $\int hg_2g_1^{-1}S$. For
\begin{equation}\label{diffhpersist}
\left (\frac{hg_2}{g_1g}\right )^\prime \frac{g_1}{hg_2}=\left ( \frac{h}{g}\right )^\prime
\frac{g_2}{g_1}\frac{g_1}{hg_2}+\frac{h}{g}\left (\frac{g_2}{g_1}\right )^\prime
\frac{g_1}{hg_2}=\left (\frac{h}{g}\right )^\prime h^{-1}+
\left (\frac{g_2}{g_1}\right )^\prime \frac{g_1}{gg_2}.
\end{equation}
Since $G_2\lt\pm\infty$ and $g_2g_1^{-1}\lt 0$, Lemma~\ref{HFsmalls_lem}(ii) gives
$(g_2g_1^{-1})^\prime =o(g_2)$. Since also $g\sim g_1$, the second term on the right of
(\ref{diffhpersist}) tends to zero. Hence the left-hand side tends to $\pm\infty$,
$0$ or an element of $\R\upstar$with $(hg^{-1})^\prime h^{-1}$, which is our above assertion.

If we are in Diff-h then (\ref{Diff-h}) and Corollary~\ref{Coroll2} yield
\[
T^{-1}\int\frac{g_2h}{g_1}\sin(G_1+G_2)=
T^{-1}\int\frac{g_2h}{g_1}S\asymp \frac{-g_2h}{Tgg_1}C\asymp g_2g_1^{-1},
\]
and similarly for $\sin(G_1-G_2)$.

For the Int-h case, (\ref{Int-h}) gives
\[
\frac{1}{T}\int\frac{hg_2}{g_1}S\asymp \frac{1}{T}\int\frac{hg_2}{g_1}
\preceq\frac{g_2\tau}{g_1},
\]
for some $\tau$ with $\gamma(\tau)<\gamma(g_2g_1^{-1})$, by Lemma~\ref{IntEqLem}.
Hence in both cases $T^{-1}$ times the first integral on the right of
(\ref{GSPL_insert}) belongs to the $\bt_i$ ghost.

As regards the second integral, suppose first that we are in Diff-h for the
inner integral. Then
\begin{eqnarray*}
T^{-1}\int\left (g_2S_2\int \left (\frac{h}{g_1}\right )^\prime C_1\right ) = &&
T^{-1}\int\frac{g_2}{g_1}\left (\frac{h}{g_1}\right )^\prime S_1S_2\\
&&  -\quad T^{-1}\int g_2S_2
\int\left (\left(\frac{h}{g_1}\right )^\prime\frac{1}{g_1}\right )^\prime S_1.
\end{eqnarray*}
The inequality $|\int F|\leq\int |F|$ and Lemma~\ref{IntEqLem} show that each
of the summands on the right has $\gamma_0$ no greater than that of
$T^{-1}g_2g_1^{-1}hg_1^{-1}\tau$ for some $\tau$ with $\gamma(\tau)<\gamma(g_2g_1^{-1})$,
and this is less than a positive real power of $\bt_i$.

If we are in the Int-h case for the second integral we have
\[
T^{-1}\int\left (g_2S_2\int \left (\frac{h}{g_1}\right )^\prime C_1\right )\preceq
\frac{1}{H}\int\frac{hg_2}{g_1}\preceq\frac{g_2\tau}{g_1},
\]
by Lemma~\ref{IntEqLem} again.

So $\int hS_1C_2 - C_2\int hS_1$ belongs to the ghost. We obtain the same conclusion
for $\int hC_1S_2-S_2\int hC_1$, and the lemma follows from (\ref{G-splitEq2}).
\end{proof}

Lemma~\ref{G_splitLem} allows us to assume that $g$ is equal to its $i$-th shadow.
If $g$ does not split as $g_1+g_2$ with $\gamma(g_2/g)\geq\gamma(\bt_i)$ and $g_2=o(g)$,
there is of course no problem. Otherwise $\sin G_2$ and $\cos G_2$ may
be taken outside the integral in $\int hS$. Note however the remarks
towards the end of Section~\ref{expn_comments}.\\

Recall that we are assuming $h\not\asymp g$. Also since we are either in the Diff-h
or the Int-h case, $h^{-1}(h/g)^\prime\not\asymp 1$; see the remarks preceding
Section~\ref{desubsec}.
We put $\alpha=h^{-1}(h/g)^\prime$ and $\beta=Hgh^{-1}$. Then
Lemma~\ref{diffintcc_lem} implies $\beta\bowtie\alpha^{-1}$. Note that
$g^{-1}T^\Delta\bowtie\alpha$ in both the Diff-h and the Int-h cases.
\begin{Def}\label{shadowsdef}
Let $A\in\R$ be equal to zero if $\gamma(T)\geq\gamma(\bt_i)$ and
equal to the arbitrary constant of the integration $\int hS$ otherwise.
The $\bt_i$-shadow of $T^{-1}\int hS$ will be shown to be $\phi_i$,
defined as follows:
\begin{enumerate}
\item If $\alpha\lt 0$ and $\gamma(\alpha)\geq\gamma(\bt_i)$, then
$\phi_i=-\eta_i(h/(Tg))C+AT^{-1}$.
\item If $\alpha\lt\infty$ and $\gamma(\alpha)\geq\gamma(\bt_i)$,
then $\phi_i=\eta_i(HT_H^{-1})S+AT^{-1}=(H_iS+A)T^{-1}$.
\item If $\gamma(\alpha),\gamma(G)<\gamma(\bt_i)$, then
$\phi_i=T^{-1}\int h_iS+AT^{-1}$.
\item If $\gamma(\alpha)<\gamma(\bt_i)\leq\gamma(G)$,
\[
\phi_i=\frac{h_i(\omega S-C)}{gT(1+\omega^2)}+\frac{A}{T},
\]
with $\omega=\eta_i(g^{-1}T^\Delta)$.
\end{enumerate}
\end{Def}
We make a number of remarks. Firstly $AT^{-1}\in\cS_i(\cA)\cap\cH$.
Secondly we note that case (ii) may necessitate the introduction
of a new scale element, \cite{Sh7,Sh_bk}. Thirdly since we are
assuming that $g\in\cS_i(\cA)$, it is immediate that $\phi_i\in\cS_i(\cA)$ in
cases (i), (ii) and (iv).

Definition~\ref{shadowsdef} again highlights the problem of the arbitrary
constant. This will be in the shadow $\phi_i$, or in the ghost
according to whether $\gamma(T)<\gamma(\bt_i)$ or not; c.f.
\cite{Sh7,Sh_bk}. For the most part we have to keep
the arbitrary constant as a parameter. It is quite possible
in case~(iii) for example, that $\cA$ or its real-algebraic closure
already contains $T^{-1}\int h_iS$ for some value of the
constant, and we might not discover this unless we are able to determine
algebraic dependencies in $\cS_i(\cA[\int hS,\phi_i])$ or we happen to test the
relevent polynomial for zero equivalence. If we
find that there is a value of the constant which does indeed make
the polynomial zero, it may well be sensible to fix the constant at that
value.

To prove the correctness of Definition~\ref{shadowsdef} we firstly
have to show that in all cases the ghost $\xi_i(T^{-1}\int hS)\in\cI_i$,
where here and in the rest of the paper $\cI_i=\cI_i(\cA[\int hS,\phi_i])$.
Secondly we have to prove that $\cS_i(\cA)(\phi_i)$ has the shadow properties
SF(i) and SF(ii).

The first requirement is met by the following.
\begin{Lem}\label{xi_lem}
In all of the above cases, $T^{-1}\int hS-\phi_i\in\cI_i$.
\end{Lem}
\par\begin{proof}%[of Lemma~\ref{xi_lem}]
In case (i) of the definition, (\ref{Diff-h}) gives
\begin{equation}\label{xiLemEq1}
\xi_i\left (\frac{1}{T}\int hS\right )=-C\xi_i\left
(\frac{h}{Tg}\right ) +\frac{1}{T}\int\left (\frac{h}{g}\right )^\prime C.
\end{equation}
Integration by parts yields
\[ \frac{1}{T}\int\left (\frac{h}{g}\right )^\prime C=
\left (\frac{h}{g}\right )^\prime\frac{S}{Tg}-\frac{1}{T}\int
\left (\left (\frac{h}{g}\right )^\prime\frac{1}{g}\right )^\prime S. \]
Here the first term on the right has $\gamma_0$ equal to that of
$\alpha=(h/g)^\prime h^{-1}$ and so belongs to $\cI_i$. Similarly for the second,
\[ \left |\frac{1}{T}\int\left (\left (\frac{h}{g}\right )^\prime
\frac{1}{g}\right )^\prime S\right |\leq\frac{1}{T}
\int\left |\left (\left (\frac{h}{g}\right )^\prime\frac{1}{g}
\right )^\prime\right |\asymp  \left (\frac{h}{g}\right )^\prime
\frac{1}{h}\in\cI_i, \]
as required.

Likewise in case (ii) we have from (\ref{Int-h})
\begin{equation}\label{xiLemEq2}
\xi_i\left (\frac{1}{T_H}\int hS\right )=
\xi_i\left (\frac{H}{T_H}\right )S-\frac{1}{T_H}\int HgC,
\end{equation}
and $\int HgC\preceq\int Hg=\int h\beta$. Moreover
Lemma~\ref{HFsmalls_lem}(viii) implies that
\[ \frac{1}{T_H}\int h\beta\asymp  \frac{\int h\beta}{\int h}
\bowtie\frac{h\beta}{h}=\beta. \]
This suffices, since $\beta\lt 0$ and $\gamma(\beta)=\gamma(\alpha)$.

Now for case (iii).
\begin{eqnarray}
\xi_i\left (\frac{1}{T}\int hS\right )&=&\frac{1}{T}\int T_h\xi_i
\left (\frac{h}{T_h}\right )S\nonumber\\
&=&-\frac{T_h}{gT}\xi_i\left (\frac{h}{T_h}\right )C+T^{-1}\int\left (
\frac{T_h}{g}\xi_i\left (\frac{h}{T_h}\right )\right )^\prime C.
\label{DfhCs2_eq1}
\end{eqnarray}
Here $T$ will either be $T_H$ or $T_hT_g^{-1}$, and these differ by a factor
of $\gamma_0$ $\gamma_0(\beta)$, which will not affect the conclusion.
Taking $T=T_hT_g^{-1}$, it is easy to see that the terms on the right
of (\ref{DfhCs2_eq1}) have $\gamma_0$ no greater than that of $\xi_i(h/T_h)$.

For the final case, let $T^{-1}\int hS=F_1S+F_2C+AT^{-1}$, where $A$ is as
in Definition~\ref{shadowsdef}.
From (\ref{f1f2-eq}), $F_k=T^{-1}f_k,k=1,2$, and (\ref{f1f2eqs}) converts to
\begin{equation}\label{F1F2eqs}
F_1^\prime=-T^\Delta F_1+gF_2+T^{-1}h,\quad F_2^\prime=-gF_1-T^\Delta F_2.
\end{equation}
We write $\chi=g^{-1}T^\Delta$; so $\omega=\eta_i(\chi)$ and we see that
$\chi\bowtie\alpha$ in each of the Diff-h and Int-h cases.
The equations (\ref{F1F2eqs}) may be solved for $F_1$ and $F_2$ in
terms of $F_1^\prime$ and $F_2^\prime$ to give
\[ F_1=\frac{1}{g(1+\chi^2)}\{\chi T^{-1}h-\chi F_1^\prime-F_2^\prime\} \]
and
\[ F_2=\frac{-1}{g(1+\chi^2)}\{T^{-1}h+\chi F_2^\prime-F_1^\prime\}. \]
We have to prove that
\[ F_1-\frac{h_i\omega}{gT(1+\omega^2)}\in\cI_i\quad \mbox{ and }\quad
F_2+\frac{h_i}{gT(1+\omega^2)}\in\cI_i. \]
We check that $g^{-1}T^{-1}(h-h_i)\in\cI_i$. Then when we substitute for $F_1$
and $F_2$ from the above equations and use $\chi-\omega\in\cI_i$, we see
that our task is a matter of showing that $g^{-1}F_k^\prime\in\cI_i$,
$k=1,2$.

If we are in Diff-h and (\ref{hseqcondition}) holds, then (\ref{h-seq}) gives
$F_1\asymp\alpha$ and $F_2\asymp  1$. Since $\gamma(\alpha)<\gamma(G)$,
Lemma~\ref{HFsmalls_lem}(iii) implies $F_1^\Delta\asymp\alpha^\Delta\prec gG^{-1}$.
Hence $g^{-1}F_1^\prime\prec F_1G^{-1}\in\cI_i$. Now $F_2^\prime$ is the derivative
of an element which tends to zero and $\log G\lt\infty$. Therefore
Lemma~\ref{HFsmalls_lem}(ii) gives
$F_2^\Delta\asymp F_2^\prime\prec G^\Delta$, and so
$g^{-1}F_2^\prime\prec F_2G^{-1}\in\cI_i$.

However we have to consider that (\ref{hseqcondition}) might fail with $j=0$,
rendering our use of (\ref{h-seq}) invalid. This can only happen if
$(h/g)^\prime\asymp g$. Then $T\asymp h/g\asymp G$. Also, with reference to
(\ref{Diff-h}), $\int(h/g)^\prime C\asymp \int gC=S$. So in this eventuality
$F_1\asymp T^{-1}\asymp G^{-1}$ and then $g^{-1}F_1^\prime\asymp G^{-2}\in\cI_i$.

In Int-h, $F_1\asymp  1$ and $F_2\bowtie\beta$ by (\ref{IntH_seq}) and
Lemma~\ref{HFsmalls_lem}, and we obtain the conclusion similarly.
\end{proof}
\subsection{The shadow property SF(i) in $\cA[\int hS]$}\label{SFi_subsec}
In cases (i), (ii) and (iv) there is nothing to prove since the shadow domain is
unchanged.

For case (iii), let
\begin{equation}\label{phi_iDef7.1}
\phi_i=\Psi_1S+\Psi_2C+AT^{-1}.
\end{equation}
Differentiation leads to the equations
\begin{equation}\label{psitoPsieq}
\Psi_1^\prime =-T^\Delta\Psi_1+g\Psi_2+T^{-1}h_i,\qquad
\Psi_2^\prime=-g\Psi_1-T^\Delta\Psi_2.
\end{equation}
c.f. (\ref{F1F2eqs}).

We shall need to use Lemma~3 from \cite{Sh7}, which we now state in a form
suitable for the present context.
\begin{Lem}\label{SFiLem}
For $j=1,2$, let $\Psi_j$ be Hardy-field elements which satisfy
differential equations
\begin{equation}\label{Psiprimeeq}
 \Psi_j^\prime =\Lambda_{j,1}\Upsilon_{j,1}^\prime+\Lambda_{j,2}
\Upsilon_{j,2}^\prime+\cdots +\Lambda_{j,k}\Upsilon_{j,k}^\prime,
\end{equation}
with each $\Lambda_{j,\nu}\in(\cS_i(\cA)\cap\cH)(\Psi_1,\Psi_2)$ and each
$\Upsilon_{j,\nu}\in\cS_i(\cA)\cap\cH$. Then $(\cS_i(\cA)\cap\cH)(\Psi_1,\Psi_2)$
satisfies SF(i).
\end{Lem}
Before we can apply Lemma~\ref{SFiLem} to our present situation we need
another lemma. 
\begin{Lem}\label{GLogTLem}
Suppose that $\gamma(\alpha)<\gamma(\bt_i)$. Then the following hold:
\begin{enumerate}
\item We have $h_ig^{-1}T^{-1}\in\cS_i(\cA)$.
\item If also $\gamma(G)<\gamma(\bt_i)$ then $G$ and $\log T$ belong
to $\cS_i(\cA)$.
\end{enumerate}
\end{Lem}
It is possible that consideration of $\log T$ might necessitate the
introduction of a new logarithm into $\cH$. However there is no difficulty
in doing this. Moreover the problem cannot reoccur with the new scale element
this brings, since that scale element would be of lowest comparability class,
and the condition $\gamma(\alpha)<\gamma(\bt_i)$ could not then be met
since $\alpha\not\asymp 1$.
\begin{proof}
For (i) suppose first that we are in Diff-h. Then $h_i/(Tg)=\eta_i(h/T_h)\cdot T_g/g$
and the only issue is whether $T_g/g$ is in $\cS_i(\cA)$. Since we are
assuming $g\in\cS_i(\cA)$ this is immediate.

In Int-h, $T=T_H$ and $g^{-1}T_H^{-1}h_i \bowtie\alpha^{-1}$ by
Lemma~\ref{diffintcc_lem}; so $g^{-1}T^{-1}h_i\in\cR_i(\cA)$. Now $h_iT_h^{-1}$
and $g^{-1}T_g$ are $\bt_i$-shadows, and so $g^{-1}T^{-1}h_i$ is the
product of the monomial $T_H^{-1}T_g^{-1}T_h$, which is in $\cR_i$, and two
$\bt_i$-shadows. Hence, using Definition~\ref{Ch5-AsymDomainDef}(ii), it is
in $\cS_i(\cA)$.

For (ii), suppose that $\gamma(G)<\gamma(\bt_i)$. All multiseries summands
of $G$ tend to $\pm\infty$ and so $G$ has a $\{\bt_1\dts\bt_{i-1}\}$
multiseries expansion. Therefore, by the remarks following
Lemma~\ref{asymdomcond_lem}, we have $G\in\cS_i(\cA)$.

We next show that $\log T\in\cR_i$.
If $\log T\asymp  G$ we already have this. Otherwise suppose first that
that we are in Diff-h so $T=T_hT_g^{-1}$.
By Lemma~\ref{HFsmalls_lem}(viii), we have $G^{-1}\log T\bowtie
T^\prime/(Tg)\asymp h^{-1}(h/g)^\prime =\alpha$.

In Int-h, $T=T_H$ and $T^\prime/(Tg)\asymp h/(gH)
= \beta^{-1}\bowtie\alpha$.
In either case we get $\log T\in\cR_i$ as desired.

But $\log T$ is an $\R$-linear combination of logarithms of
different scale elements. All of these logarithms tend to $-\infty$
and their $\gamma_0$s are all different. So each must have comparability
class less than $\gamma(\bt_i)$, which implies that $\log T\in\cS_i(\cA)$.
\end{proof}

Lemma~\ref{GLogTLem} shows that the equations (\ref{psitoPsieq}) are
in the form of (\ref{Psiprimeeq}) with the various $\Upsilon$ being
either $G$ or $\log T$.
Lemma~\ref{SFiLem} then implies SF(i) for
the case $\gamma(G),\gamma(\log T)<\gamma(\bt_i)$.
So we have proved the following.
\begin{Lem}\label{SFi_Done}
The shadow property SF(i) holds in $(\cS_i(\cA)\cap\cH)(\Psi_1,\Psi_2)$.
\end{Lem}
\subsection{The Shadow property SF(ii)}\label{shadowSFii}
We may again assume that we are in Case~(iii) of
Definition~\ref{shadowsdef}. We have to show that if $P$ is
a polynomial over the domain $\cS_i(\cA)$ and
$P(\phi_i)\in\cI_i$ then $P(\phi_i)=0$. Write
\[ P(Z)=P_DZ^D+P_{D-1}Z^{D-1}+\cdots+P_0, \]
with $P_0\dts P_D\in\cS_i(\cA)$.
We assume that $D$ is minimal for such polynomials.
Asymptotic independence is then used to establish the following.
\begin{Lem}\label{SandCLem}
If there is a non-zero polynomial, $P$, over $\cS_i(\cA)$ such that
$P(\phi_i)\in\cI_i$, then there is such a polynomial whose
coefficients belong to $(\cS_i(\cA)\cap\cH)[S,C]$.
\end{Lem}
\begin{proof}%[of Lemma~\ref{SandCLem}]
Suppose  that $s$ is a sine or cosine different from $S$ and $C$
which appears in the coefficients of $P$. Then we may write
$P(Z)=s\widehat{P}(Z)+\widetilde{P}(Z)$ with the coefficients of
$\widetilde{P}$ not containing $s$.

We expand out the powers of $S$ and $C$ in $\widehat{P}(\Psi_1S+\Psi_2C)$
and $\widetilde{P}(\Psi_1S+\Psi_2C)$. It is then clear from
Lemma~\ref{indep_prods_lem} that $\widetilde{P}(\phi_i)\in\cI_i$.
Therefore we may replace $P$ by $\widetilde{P}$ unless $\widetilde{P}$
is the zero polynomial. If it is, then $s$ is a factor of $P$ which
we may cancel out. If we remove all such factors in this way,
we obtain Lemma~\ref{SandCLem}.
\end{proof}

We henceforth assume that $P$ is as given by the conclusion of this lemma.
Our basic idea for proving that $P(\phi_i)\in\cI_i$ implies $P(\phi_i)=0$
is to differentiate $P(\phi_i)$ and substitute for $\phi_i^\prime$ from its
differential equation, $\phi_i^\prime = T^{-1}h_iS-T^\Delta \phi_i$.
We let 
\[ Q(Z)=\partial_{Z}P(Z)\{T^{-1}h_iS-T^\Delta Z\}+\partial_cP(Z), \]
where $\partial_c$ indicates differentiation of the coefficients,
and check that $P(\phi_i)^\prime=Q(\phi_i)$.
\begin{Lem}\label{subcase2aLem}
The coefficients of $g^{-1}Q(Z)$ belong to $\cS_i(\cA)$.
\end{Lem}
\par\begin{proof}%[of Lemma~\ref{subcase2aLem}]
For $g^{-1}T^{-1}h_i$ this is just Lemma~\ref{GLogTLem}(i).
Now SF(i) in $\cS_i(\cA)$ and Lemma~\ref{GLogTLem}(ii) show that
$T^\Delta/g\in\cS_i(\cA)\cap\cH$. Similarly SF(i) implies that the
coefficients of $g^{-1}\partial_cP(Z)$ belong to $\cS_i(\cA)$.
\end{proof}

Now
\begin{equation}\label{PQSubtract}
g^{-1}P_DQ(Z)+(DT^\Delta P_D-P_D^\prime)g^{-1}P(Z)
\end{equation}
is easily seen to be a polynomial over $\cS_i(\cA)$. The terms in $Z^D$
cancel, so it is of degree at most $D-1$. However
Lemma~\ref{HFsmalls_lem}(viii) tells us that $G^{-1}P(\phi_i)\bowtie
g^{-1}P(\phi_i)^\prime$ and so the latter belongs to $\cI_i$.
Thus (\ref{PQSubtract}) produces an element of
$\cI_i$ when the substitution $Z=\phi_i$ is made, and the minimality of $D$
now forces it to vanish identically when $Z=\phi_i$.

If $P(\phi_i)\neq 0$, we may divide (\ref{PQSubtract}) through by $g^{-1}P_DP(\phi_i)$
to give the identity $P(\phi_i)^\Delta = P_D^\Delta-DT^\Delta = (P_DT^{-D})^\Delta$.
As in the proof of Theorem~\ref{RatFnsTh}, we may  integrate this identity
on the sequence of intervals $\{I_\nu\}$ on which $P_DP(\phi_i)$ has no zeros,
to give
\begin{equation}\label{Sec7.2PphiIdentity}
P(\phi_i)=KP_DT^{-D}
\end{equation}
with $K$ non-zero and constant on each $I_\nu$.

Theorem~\ref{int_flutter_th} allows us to apply Lemma~\ref{zerolts_lem},
and so these intervals abut. Let $p$ be the common endpoint of $I_\nu$
and $I_{\nu+1}$ for some $\nu$; we may take $p$ to be sufficiently
large. If $P(\phi_i)\neq 0$
some derivative $P(\phi_i)^{(k)}$ must be non zero at $p$, and we let
$k\in\Z^+$ be the least number for which this is so. Continuity of
$P(\phi_i)^{(k)}$ at $p$ forces the values of $K$ to be the same in
$I_\nu$ and $I_{\nu+1}$. Hence (\ref{Sec7.2PphiIdentity}) holds with
$K\in\R\upstar$.

If $\gamma(T)<\gamma(\bt_i)$ or if $T\lt 0$ we have a contradiction to the
assertion that $P(\phi_i)\in\cI_i$, and so $P(\phi_i)$ must have been zero
all along.

Otherwise $\gamma(T)\geq\gamma(\bt_i)$ and $T\lt\infty$. We take
$\lambda$ to be a real parameter and
write $\psi_\lambda=\phi_i+\lambda/T$. Then $\psi_\lambda$ satisfies the same
differential equation as $\phi_i$ and $P(\psi_\lambda)\in\cI_i$. Therefore the
working above may be carried out with $\psi_\lambda$ replacing $\phi_i$, giving
$P(\psi_\lambda)=T^{-D}U(\lambda)P_D$.
From the form of (\ref{Sec7.2PphiIdentity}) it is clear that $U(\lambda)$
is a polynomial over $\R$ of degree no more than $D$.
Let $U(\lambda)=u_D\lambda^D+u_{D-1}\lambda^{D-1}+\cdots +u_0$.
Comparison of coefficients of $\lambda^{D-1}$ gives $DP_D\phi_i+P_{D-1}= u_{D-1}P_D/T$,
which belongs to $\cI_i$. But
the coefficients of $DP_D\phi_i+P_{D-1}$ belong to $\cS_i(\cA)$, and so the minimality
of $D$ now requires $D=1$. Thus, reverting to the case $\lambda=0$,
\begin{equation}\label{neqoneeq}
P_1\phi_i+P_0=KT^{-1},\quad K\in\R\upstar
\end{equation}
We note that in the case we are considering, the constant $A$ in (\ref{phi_iDef7.1})
must be zero. The idea then is to substitute $\phi_i=\Psi_1S+\Psi_2C$ with $\Psi_1$
and $\Psi_2$ Hardy-field elements and compare
coefficients of the various powers of $S$ and $C$. We assume $P_0$ and $P_1$
contain no powers of $C$ greater than one, and we let $N$ be the total degree
of $S$ and $C$ in $P_1$. If $N=0$ then (\ref{neqoneeq}) implies
that the coefficient of $S^0C^0$ in $P_0$ is equal to  $K/T$, which is impossible
since $P_0\in\cS_i(\cA)$ and $K\neq 0$.

So suppose $N\geq 1$. We write $\Omega_0$ and $\Theta_0$ for the coefficients
of $S^{N+1}$ and $CS^N$ in $P_0$ and $\Omega_1,\Theta_1$ for the coefficients
of $S^N$ and $CS^{N-1}$ in $P_1$ respectively. Terms of total degree $N+1$
are
\[ (\Omega_1S^N+\Theta_1CS^{N-1})(\Psi_1S+\Psi_2C)+\Omega_0S^{N+1}+\Theta_0CS^N, \]
whence
\begin{equation}\label{SNplusoneEq}
\Omega_1\Psi_1-\Theta_1\Psi_2+\Omega_0=0,\qquad\mbox{and}
\qquad \Theta_1\Psi_1+\Omega_1\Psi_2+\Theta_0=0. 
\end{equation}
The discriminant of this pair is $\Omega_1^2+\Theta_1^2$, which is non-zero
by defininition of $N$.
Therefore (\ref{SNplusoneEq}) may be solved to give $\Psi_1$ and $\Psi_2$ as
elements of $\cS_i(\cA)$. But then $\phi_i\in\cS_i(\cA)$ and this gives SF(ii).
We have therefore proved the following.
\begin{Th}\label{SFiiThm}
If $\phi_i=\eta_i(T^{-1}\int hS)$ is as in Definition~\ref{shadowsdef},
then $\cS_i(\cA)(\phi_i)$ is a $\bt_i$- shadow domain of $\cA[\int hS]$.
\end{Th}
\subsection{Shadows of shadows and of ghosts}
In fact $\eta_j(\phi_i), j<i$ poses little difficulty. For in cases~(i), (ii)
and (iv) of Definition~\ref{shadowsdef}, $\phi_i\in\cS_i(\cA)$, while in case~(iii)
we apply Definition~\ref{shadowsdef} with $h$ replaced by $h_i$ and $i$ by $j$.
The map $\eta_{i,j}$ is then defined on $\cS_i(\cA)(\phi_i)$ in the obvious way.

In case (iii) we obtain $\eta_j(\xi_i(T^{-1}\int hS))$, $j\leq i$, in similar
fashion to $\eta_j(\phi_i)$. For other cases more work is needed though.
We begin with an independence result for integrals.
\begin{Lem}\label{IntIndepLem}
Let $h,\widetilde{h}$ be positive elements of $\cH$. Then
\[
\gamma_0\left (\int hS+\int\widetilde{h}C\right )=\max\left\{\gamma_0\left (\int hS
\right ),\gamma_0\left (\int\widetilde{h}C\right )\right \}.
\]
\end{Lem}
\par\begin{proof}%[of Lemma~\ref{IntIndepLem}]
Unless $\widetilde{h}\asymp h$ this is immediate from the corollaries to
Theorem~\ref{int_flutter_th}. Suppose then that $\widetilde{h}=Kh+\varepsilon$ with
$K\in\R\upstar$ and $\varepsilon=o(h)$. As in the proof of Theorem~\ref{RatFnsTh},
it will be sufficient to prove the result when $\varepsilon=0$.

Recall that $\int hS=f_1S+f_2C$ and $\int hC=f_1C-f_2S$. Then
\[ \int hS+\int\widetilde{h}C=(f_1-Kf_2)S+(f_2+Kf_1)C. \]
Lemma~\ref{indep_prods_lem} shows that if there is to be cancellation, we must
have $f_1\sim Kf_2$ and $f_2\sim -Kf_1$. But this implies $f_1\sim -K^2f_1$,
which is only possible if $f_1=0$. Similarly $f_2=0$, which is a clear
contradiction since we have assumed that $h>0$.
\end{proof}

Let $\xi_i=\xi_i(T^{-1}\int hS)$. An inspection of the proof of Lemma~\ref{xi_lem},
shows that in case (i) we can write $\xi_i$ in the form $F+T^{-1}\int\zeta C$,
with $\zeta\in\cH$ and $F,\:T^{-1}\int\zeta S,\:T^{-1}\int \zeta C\in\cI_i(\cA[\int hS,\phi_i])$.
Then writing $F$ as $T^{-1}\int (TF)^\prime$ gets $\xi_i$ into the form
\begin{equation}\label{xiSCeq}
\xi_i=T^{-1}\left \{\int\zeta_1S+\int\zeta_2C\right \}
\end{equation}
with $\zeta_1,\zeta_2\in\cH$. Case~(ii) is very similar.

In case~(iv) the definition similarly allows us to write $\xi_i$ in this form.
Lemma~\ref{IntIndepLem} assures us that the leading terms of the two integrals
in (\ref{xiSCeq}) cannot cancel, and we can therefore compute their
$\bt_j$-shadows separately. This we are able to do by the above methods.
It is clear that we can continue the expansion in this way.

In order to obtain the form (\ref{xiSCeq}), it may in practice be more
efficient in cases~(i) and (ii) to use (\ref{xiLemEq1})
and (\ref{xiLemEq2}) respectively, rather than the definition directly.

It is now clear from Definition~\ref{shadowsdef} and the corresponding
relations in $\cH$ that $\eta_{j,i}\circ\eta_i=\eta_j$, for $1\leq j<i\leq m$,
and $\eta_{k,j}\circ\eta_{j,i}=\eta_{k,i}$ for $1\leq k<j<i\leq m$.

To integrate a general element of $\cA$, we can use the trigonometric
addition formulae to write it in the form $\sum_j(h_{1,j}\sin G_j+
h_{2,j}\cos G_j)$, although some shortcuts may be provided by
Lemma~\ref{G_splitLem}.

We have thus proved the following.
\begin{Th}\label{AintAAsFld}
Suppose that $\cA=\cA$. Then $\cA[\int hS]$ is an asymptotic domain.
\end{Th}
\subsection{Some comments about the expansion of $\int hS$}\label{expn_comments}
It should be noted that substituting the multiseries for the various $h_{\langle n\rangle}$ or
$H_{\langle n\rangle}$ into (\ref{h-seq}), respectively (\ref{IntH_seq}), is not always correct.
The problem is indefinite cancellation, \cite{Sh1,Richardsonetal,Gruntz96,Sh_bk}.
The following example illustrates the point, although there are probably simpler
ones and also better hidden ones. Let
\begin{equation}\label{indefExample}
F=\int \frac{\sin(\log x)}{x\log x+\sqrt{x}}+\frac{\cos(\log x)}{\log x}
+\int\frac{\cos(\log x)}{x\log^2x}.
\end{equation}
An induction shows that for the first integral 
\[ h_{\langle n\rangle} =\frac{(-1)^nn!(1+\varepsilon_n)}{x\log^{n+1}x}, \]
with $h_{\langle n\rangle}$ as in (\ref{h-seq}) and $\varepsilon_n\preceq x^{-1/2}$
for all $n$. The leading terms of the multiseries for the first integral
are therefore
\[ \sum (-1)^{n+1}\left\{\frac{(2n)!\cos(\log x)}{\log^{2n+1}x}+
\frac{(2n+1)!\sin(\log x)}{\log^{2n+2}x}\right \}. \]
These cancel term by term with those from the other summands in
(\ref{indefExample}) and the calculation will fail to terminate.
The terms of comparability class $\gamma(x^{-1})$ will never be reached.
However testing of the $x^{-1}$-shadow of $F$ reveals that it to be zero.
So we then know that expansion has to begin with the $x^{-1}$-ghost.
Thus the methods described above will produce the correct multiseries.

Care is required when the expression to be expanded contains more than one
integral. For example it is easy to show that for $\int\sin x^k$, $k\in\R^+$,
the terms of (\ref{h-seq}) decrease by $O(x^k)$. So for
$\int\sin x^2+\int\sin x^{2/3}$ these interlace.

As a spin-off from the results about $\int hS$, we can generate multiseries
expansions for solutions of the second-order differential equation
(\ref{f2diff-eq}). For $f_2=C\int hS-S\int hC$ is a solution 
and others are obtained as $f_2+K_1S+K_2C$ with $K_1,K_2\in\R$.

We can replace $\cA$ by $\cA[\int hS]$ above and compute multiseries
for new integrals there; then we can repeat the process. Also
the function $f_2$ itself might form part of a trigonometric argument.
So we could generate multiseries for $\int h\sin (hf_2)$ for example.

If we want to compute the $\bt_i$-shadow of $T^{-1}\int h\sin G$, where
$G=(G_1+G_2)$ and we are in the situation where Lemma~\ref{G_splitLem}
applies, it might seem that we need to use the addition formula and
process $T^{-1}\int hS_1$ and $T^{-1}\int hC_1$ separately, but in practice
we don't.

With $S_1,C_1$ etc. as in Lemma~\ref{G_splitLem}, let $\Psi_1S_1+\Psi_2C_1$
and $\Psi_1C_1-\Psi_2S_1$ be the $\bt_i$-shadows of
$T^{-1}\int hS_1$ and $T^{-1}\int hC_1$ respectively. Then the lemma gives
\begin{eqnarray*}
\eta_i\left (\frac{1}{T}\int h\sin G\right )&=&
C_2\{\Psi_1S_1+\Psi_2C_1\}+S_2\{\Psi_1C_1-\Psi_2S_1\}\\
&=&\Psi_1\sin(G_1+G_2)+\Psi_2\cos(G_1+G_2).
\end{eqnarray*}
This is exactly the result we get by not expanding $\sin(G_1+G_2)$
but replacing $G^\prime$ with $\delta_i(G)$, where $\delta_i$ is
the derivation $\eta_i\circ d/dx$. In other words, we
can throw away the `tail' of $g$ while retaining $G$ exactly as it is.

\end{document}